\documentclass[DIV=15,11pt,abstract]{scrartcl}
\usepackage[utf8]{inputenc}
\usepackage{amssymb, mathtools, bm, amsthm, thmtools, relsize}
\usepackage{float}
\usepackage{graphicx} 
\usepackage{authblk}
\usepackage[bottom]{footmisc}   

\usepackage[toc,page]{appendix}
\usepackage[noadjust]{cite} 
\usepackage{xcolor}
\definecolor{myBlue}{RGB}{1,1,141}

\usepackage[colorlinks=True]{hyperref}
\hypersetup{allcolors=myBlue}

\newcommand{\cB}{\mathcal{B}}
\newcommand{\cC}{\mathcal{C}}

\newcommand{\cG}{\mathcal{G}}

\newcommand{\cJ}{\mathcal{J}}

\newcommand{\cS}{\mathcal{S}}

\newcommand{\ba}{\mathbf{a}}
\newcommand{\bb}{\mathbf{b}}
\newcommand{\bc}{\mathbf{c}}
\newcommand{\bd}{\mathbf{d}}
\newcommand{\be}{\mathbf{e}}
\newcommand{\bf}{\mathbf{f}}

\title{Large-$N$ limit of $O(N)^3$-invariant general sextic tensor model}

\author[1]{G.~Bardy}
\author[2]{T.~Krajewski} 
\author[3]{T.~Muller}
\author[1,4]{A.~Tanasa}

\affil[1]{\textit{Univ. Bordeaux, LaBRI CNRS UMR 5800, Talence, France}}
\affil[2]{\textit{Aix Marseille Univ, Université de Toulon, CNRS, CPT, Marseille, France}}
\affil[3]{\textit{Univ. Sorbonne Paris Nord, LIPN UMR CNRS 7030, Villetaneuse, France}}
\affil[4]{\textit{DFT, H. Hulubei Nat. Inst. Phys. Nucl. Engineering, Magurele,  Magurele, Romania}}

\begin{document}


\maketitle

\begin{abstract}
We study a sextic tensor model where the interaction terms are given by all $O(N)^3$-invariant bubbles. The class of invariants studied here is thus a larger one that the class of the 
$U(N)^3$-invariant sextic tensor model. 
We implement the large $N$ limit mechanism for our model and we 
explicitly identify the dominant graphs in the $1/N$ expansion.
This class of dominant graphs contains tadpole graphs, melonic graphs but also new types of tensor graphs.
Our analysis adapts the tensorial intermediate field method, previously applied only to
the prismatic interaction, to all connected sextic interactions except the wheel
interaction, which we treat separately using a cycle analysis. 

\end{abstract}

\section{Introduction}

The large-$N$ limit~\cite{Moshe_2003} is an important tool in quantum field theory. 
By extending the number of degrees of freedom $N \to \infty$ and suitably  rescaling  the coupling constants by powers of $N$, one reorganizes 
the perturbative expansion in powers of $1/N$,
often allowing for controlled, analytic computations in 
otherwise intractable systems. 

We know so far three type of QFT models where one can successfully implements the large $N$ limit mechanism: 
vector, matrix, and tensor models.
In vector models~\cite{Eyal_1996}, 
cactus diagrams dominate the $1/N$ expansion, rendering the theory essentially solvable. 
Such models are central to the study of critical phenomena. 
Matrix models generate ribbon graphs (or combinatorial maps) \textit{via} perturbative expansion, their large-$N$ limit is dominated by planar 
diagrams~\cite{HOOFT1974461, Brezin, KAZAKOV1985295, DAVID198545}. 
 The sum over ribbon graphs, organized by the genus,  corresponds to a sum over 
discretized two-dimensional surfaces thus providing a natural framework 
for random geometry in $d=2$~\cite{marino2005leshoucheslecturesmatrix, eynard2018randommatrices} 
and for two-dimensional quantum gravity~\cite{ginsparg1991matrix, ginsparg1993lectures2dgravity2d, 
difrancesco20042dquantumgravitymatrix}. 

Tensor models were 
originally introduced as descriptions of random geometries in higher dimension
\cite{sasakura_tensor_1991, ambjorn_three-dimensional_1991}. 
Their large-$N$ behavior was
discovered much later, see the books ~\cite{10.1093/acprof:oso/9780198787938.001.0001, 
10.1093/oso/9780192895493.001.0001}. It is 
is governed by melonic diagrams~\cite{Gurau_2011}, which 
form a subset of planar diagrams. The melonic limit is, from this diagrammatic point of view, simpler than the planar one 
yet richer than the cactus one, and thus captures nontrivial dynamics. 

The same melonic large-$N$ limit appears in the celebrated Sachdev--Ye--Kitaev (SYK) model, 
initially introduced for spin glasses~\cite{Sachdev_1993}. The SYK model describes 
$N$ Majorana fermions with random interactions, averaged over disorder, and has 
attracted a wide interest as a toy model for holography~\cite{Kitaev_2015}. Tensor 
models provide a disorder-free reformulation of the SYK model~\cite{witten2016syklikemodeldisorder, 
GURAU2017386, Krishnan_2017}, motivating the development of tensor field theories in $d>0$~\cite{Giombi_2017, Giombi_2018, Benedetti_2019, jepsen2023rg}.

The $1/N$ expansion of tensor models is controlled by the degree 
\cite{Gurau_2011, gurau_complete_2012}. 
Understanding the large-$N$ limit of a specific model is a prerequisite both for 
studying it as a field theory and for combinatorial analyses such as the double 
scaling limit \cite{Bonzom_2022, Krajewski_2023}. This paper addresses this question 
for the $O(N)^3$-invariant sextic tensor model. 

Recall that $O(N)^3$-invariant tensor models were first introduced in \cite{Carrozza_2016}, 
where the large-$N$ limit of the quartic model was implemented. Field-theoretic 
studies of this model followed in \cite{Giombi_2017, Benedetti_2019}. In the sextic 
case, the large-$N$ limit of the $U(N)^3$ model was derived in \cite{bonzom2015colored} (see also \cite{LIONNI2019600, Lionni_2018}). Its set of interactions is a subset of those of the $O(N)^3-$invariant model that we study in this paper.  A renormalization group study for the $U(N)^3-$invariant model was then performed in \cite{Benedetti_2020}. 

The so-called
prismatic model \cite{Giombi_2018} was also solved at large $N$, though with 
scalings which favor a particular sextic interaction, the prismatic one (this is motivated by the renormalization group study performed in that paper). 
A sextic model with a reduced set of $O(N)^3$-invariant
interaction was also studied in \cite{Prakash_2020}. More 
recently, perturbative field-theoretic computations on the full $O(N)^3$ 
sextic model, including a Yukawa interaction, were carried out 
in~\cite{Fraser-Taliente:2024rql}.  

Nevertheless, the large-$N$ limit of the  sextic model 
taking into consideration all $O(N)^3$ invariant interactions
with optimal scalings has remained 
unknown. In this paper, we implement 
this large $N$ limit 
and we explicitly exhibit the dominant graphs. 

The class of dominant graphs we find is significantly larger than the class of graphs found for the prismatic model  \cite{Giombi_2018} or the one found for
the $U(N)^3$ invariant model. This is a direct consequence of the fact that the class of $O(N)^3$ invariant interaction is  larger than the class of $U(N)^3$ invariant interactions.

Let us emphasize that this class of dominant graphs we find here contains tadpole graphs, the celebrated melonic graphs but also new types of graphs.

To identify these dominant Feynman graphs, we employ the intermediate field method 
\cite{1957SPhD....2..416S, PhysRevLett.3.77}, which is adapted in a non-trivial way to tensorial interactions, see  
\cite{Rivasseau_2013, lionni2016intermediate, Krajewski_2023}. 
In order to obtain the class of dominant graphs of our model, we prove that 
the intermediate field method decomposes most of the 
sextic tensor interactions into quartic ones and it reduces, from a diagrammatic point of view,
the analysis 
of our general model
to a minimal set of 
interactions: the tetrahedron and the wheel $K_{3,3}$.

The paper is organized as follows.
Section~\ref{Sec2} introduces the
$O(N)^3$-invariant sextic model, which has $8$ invariant bubbles.
Moreover, we define the notion of $l$-cycles in Feynman 
graphs. 
Section~\ref{Sec3} exhibits the intermediate field method for all connected sextic interactions,
except the wheel interaction.
We also prove in section \ref{Sec4} that, from a diagrammatic point of view, the general sextic model is equivalent to a reduced model containing only wheel and tetrahedron interaction. Section~\ref{Sec4} then analyzes this reduced model.  
We bound the number of cycles in 
dominant graphs, and we show
that no fundamental 
dominant graph can contain  a wheel interaction and another sextic interaction. 
Section~\ref{Sec5} 
then gives the lists of dominant graphs of our general model.
We also explicitly exhibit the $2$-point and $6-$point insertions which conserve the degree.
The last section is dedicated to the conclusions of the paper.

\section{The general sextic model}
\label{Sec2}
\subsection{The action of the model}

Random tensor models 
are $0-$dimensional field theoretical models, where the fields are
rank $r$ tensors $T^{i_1i_2\dots i_r}$, with  $i_1, \ldots i_r \in \{ 1, \ldots , N\}$. 


One usually imposes invariance of tensor models under the action of $r$ copies of a certain group $G$ (see again the books \cite{10.1093/acprof:oso/9780198787938.001.0001, 10.1093/oso/9780192895493.001.0001}). 
The tensor transforms as
\begin{equation}
    T_{a_1a_2\dots a_r}\rightarrow R^{(1)}_{a_1b_1}R^{(2)}_{a_2b_2}\cdots R^{(r)}_{a_rb_r}T_{b_1b_2\cdots b_r}, 
    \label{Gtransfo}
\end{equation}
where the $R^{(k)}$ are in the fundamental representation of $G$. The action of the given model has to be invariant under the transformation (\ref{Gtransfo}).
 The group $G$ can thus be $O(N)$ (see \cite{Carrozza_2016, Giombi_2018, jepsen2023rg}), $U(N)$ \cite{Benedetti_2020, harribey_sextic_2022} or $Sp(2N)$ \cite{Carrozza:2018psc}. Note that the duality between $O(N)$ and $Sp(2N)$ invariance \cite{mkrtchyan_equivalence_1981}, known for vector and matrix models \cite{mulase_duality_2003}, also holds in the case of random tensors (see \cite{gurau2022dualityorthogonalsymplecticrandom, Keppler_2023_1, Keppler_2023_2, TheseTM}). \\

The partition function 
of such a model writes

\begin{equation}
    \mathcal{Z}= \int[\text{d}T] e^{-N^{r/2}S[T]}.
\end{equation}
where the action writes: $$S[T]=S_0[T]+S_\text{int}[T].$$
The free part writes: $$S_0[T]=\frac{1}{2}T_{i_1,\dots,i_r}T_{i_1,\dots,i_r}.$$
The interacting part of the action is a sum over the so-called bubbles (which are the interacting terms)
\begin{equation}
    S_\text{int}[T]=\sum_b \frac{g_b}{N^{\rho(b)}} I_b(T),
    \label{genericAction}
\end{equation}
where $I_b(T)$ is invariant under the action of $G^r$ and $\rho$  is an appropriate scaling of the bubble. \\

The partition function for the general $O(N)^3$-invariant sextic model writes
\begin{equation}
    \mathcal{Z}=\int[\text{d}T]e^{-N^{3/2}S[T]},
\end{equation}
with
\begin{equation}
   S[T]=\frac{1}{2}T_{i_1i_2i_3}T_{i_1i_2i_3}+ \sum_{b=1}^8 
    \frac{g_b N^{-\rho(b)}}{6} I_b(T),
    \label{action}
\end{equation}
where $I_b(T)$ denotes the $b$-th invariant bubble and $\rho(b)$ its scaling.  

\begin{figure}[H]
    \centering
    \includegraphics[width=\textwidth]{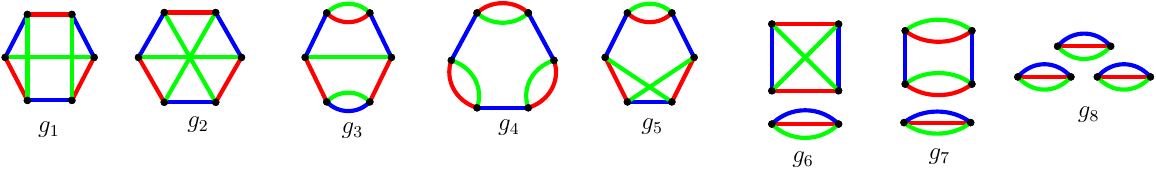}
    \caption{The $O(N)^3$-invariant sextic interactions. Color permutations are implicit.}
    \label{fig:interactions}
\end{figure}

The sextic bubbles have the general form
\begin{equation}
    I_b(T)=\delta^{(b)}_{\ba\bb\bc\bd\be\bf}T_\ba T_\bb T_\bc T_\bd T_\be T_\bf, 
\end{equation}
where $\ba=(a_1a_2a_3)$ and $\delta_{\ba\bb}=\prod_{i=1}^3\delta_{a_ib_i}$. The indices contractions, symmetrized under color permutations, but not under external indices, are
\begin{equation}
    \begin{split}
        \delta^{(1)}_{\ba\bb\bc\bd\be\bf}&=\delta_{a_1b_1}\delta_{b_2c_2}\delta_{c_1d_1}\delta_{d_2e_2}\delta_{e_1f_1}\delta_{f_2a_2}\delta_{a_3e_3}\delta_{b_3d_3}\delta_{f_3c_3},\\
        \delta^{(2)}_{\ba\bb\bc\bd\be\bf}&=\delta_{a_1b_1}\delta_{b_2c_2}\delta_{c_1d_1}\delta_{d_2e_2}\delta_{e_1f_1}\delta_{a_2f_2}\delta_{a_3d_3}\delta_{b_3e_3}\delta_{c_3f_3},\\
        \delta^{(3)}_{\ba\bb\bc\bd\be\bf}&=\frac{1}{3}(\delta_{a_1f_1}\delta_{a_2b_2}\delta_{a_3b_3}\delta_{b_1c_1}\delta_{c_2f_2}\delta_{c_3d_3}\delta_{e_3f_3}\delta_{d_1e_1}\delta_{d_2e_2}+(1\leftrightarrow 2)+(1\leftrightarrow 3)),\\
        \delta^{(4)}_{\ba\bb\bc\bd\be\bf}&=\frac{1}{3}\sum_{i=1}^3\delta_{a_ib_i}\delta_{c_id_i}\delta_{e_if_i}\prod_{j\neq i}\delta_{b_jc_j}\delta_{d_je_j}\delta_{a_jf_j},\\
        \delta^{(5)}_{\ba\bb\bc\bd\be\bf}&=\frac{1}{3}(\delta_{a_1f_1}\delta_{a_2b_2}\delta_{a_3b_3}\delta_{b_1c_1}\delta_{c_2e_2}\delta_{c_3d_3}\delta_{d_1e_1}\delta_{e_3f_3}\delta_{d_2e_2}+(1\leftrightarrow2)+(1\leftrightarrow3)),\\
        \delta^{(6)}_{\ba\bb\bc\bd\be\bf}&=\delta_{a_1b_1}\delta_{a_2c_2}\delta_{a_3d_3}\delta_{b_2d_2}\delta_{b_3c_3}\delta_{c_1d_1}\delta_{\be\bf},\\
          \delta^{(7)}_{\ba\bb\bc\bd\be\bf}&=\frac{1}{3}(\delta_{a_1b_1}\delta_{a_2b_2}\delta_{b_3c_3}\delta_{c_1d_1}\delta_{c_2d_2}\delta_{a_3d_3}+(1\leftrightarrow2)+(1\leftrightarrow3))\delta_{\bf\be},\\   
        \delta^{(8)}_{\ba\bb\bc\bd\be\bf}&=\delta_{\ba\bb}\delta_{\bc\bd}\delta_{\be\bf}.
    \end{split}
\end{equation}
The interaction terms are represented on Fig. \ref{fig:interactions}. Note that we call $I_1$ the prismatic interaction and $I_2$ the wheel interaction.

In~\cite{Giombi_2018} the scalings were chosen to be
\begin{equation}
    \rho_1 = 0, \quad 
    \rho_2=\rho_4=\rho_6=\rho_7 = 2, \quad 
    \rho_3=\rho_5 = 1, \quad 
    \rho_8 = 4.
\end{equation}

As already mentioned in the introduction, this choice of scaling was made in order to favor the prismatic interaction, because it is for this interaction that a thorough RG flow analysis was performed. 
However, we chose here a different set of scalings, which, following \cite{Carrozza_2016}, is an optimal one, in the sense that all interactions can contribute in the large $N$ limit.
These scalings are
\begin{equation}
    \rho(b) = \tfrac{1}{2} \sum_l \delta_l^{(b)},
\end{equation}
where $\delta_l^{(b)} = |J_l^{(b)}| - 1$, and $|J_l^{(b)}|$ is the number of 
connected components of the $l$-th jacket of the bubble $b$, which is obtained by removing all the edges of color $l$ in the bubble $b$ \cite{Carrozza_2016}. 


For each sextic bubble we compute $\rho(b)$ explicitly, obtaining the scalings shown in Fig.~\ref{fig:optimal scaling} (see Appendix ~\ref{optimal scalings}).  
One finds:
\begin{equation}
    \rho_1=\rho_2 = 0, \quad 
    \rho_3=\rho_4 = 1, \quad 
    \rho_5 = \tfrac{1}{2}, \quad 
    \rho_6 = \tfrac{3}{2}, \quad 
    \rho_7 = 2, \quad 
    \rho_8 = 3.
\end{equation}

Note that, for the $U(N)^3$-invariant interactions 
$I_2, I_3, I_4, I_7, I_8$, our choice of scalings 
is the same as the one of 
~\cite{Benedetti_2020}. 
Recall that the 1/$N$ expansion of tensor models is controlled by the degree of the graphs. By definition, the dominant graphs in the large $N$ limit have vanishing degree \cite{Gurau_2011}. The degree of a graph $\cG$ in this model is (see again \cite{Carrozza_2016} for the general formula)

\begin{equation}
    \omega(\cG)=3+\sum_{b\in V(\cG)}\Big(3+\rho(b)\Big)-F(\cG),
\label{omega6}
\end{equation}
where $V(\cG)$ is the set of interactions of the graph and $F(\cG)$ is the number of faces of the graph.

\subsection{Inequivalent cycles}

The analysis of the $O(N)^3$-invariant sextic model requires a careful study 
of cycles in Feynman graphs. We collect here some useful definitions and 
results, following the presentation in~\cite{Prakash_2020}.

A \emph{cycle of length $l$}, or \emph{$l$-cycle}, is a set of edges visiting 
$l$ distinct bubbles,
\begin{equation}
    \cC_l = \langle (V'_1, V_2), (V'_2, V_3), \dots, (V'_{l-1}, V_l), 
    (V'_l, V_1)\rangle,
\end{equation}
where $V_i \neq V'_i$ are vertices in the $i$-th bubble graph (not vertices 
of the full Feynman graph).

\begin{figure}[H]
    \centering
    \includegraphics[width=0.8\textwidth]{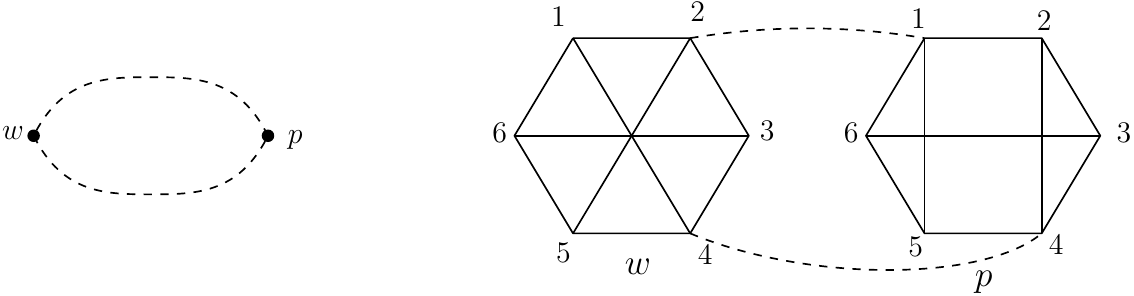}
    \caption{Example of a 2-cycle visiting a wheel and a prism in the Feynman (left) and bubble (right) representation, 
    $\cC_2 = \langle (w_2, p_1), (p_4, w_4)\rangle$.}
    \label{fig:cycle_exemple}
\end{figure}

Let us note that a cycle is not necessarily a closed loop in the bubble representation (see Fig. \ref{fig:cycle_exemple}).
When considering Feynman graphs where the internal structure of bubbles is not taken into consideration, the notion above boils down to the standard notion of a cycle in graph theory.\\

For a configuration of $l$ bubbles, let $z_i$ denote the number of vertices 
in the $i$-th bubble. The total number of $l$-cycles is then $N_{\cC_l} = \prod_{i=1}^l z_i(z_{i}-1)$.
For example, in the sextic model there are $900$ distinct $2$-cycles.  

Many of these cycles are equivalent under the symmetry group of the bubbles.  
Each bubble is invariant under two types of transformations: color 
permutations and automorphisms of the bubble. This defines a symmetry 
group $\cS_\cG$, acting as permutations on the vertex set of the full graph 
$\cG$.

Two $l$-cycles $\cC_l$ and $\cC'_l$ are \emph{equivalent} if there exists 
$\sigma \in \cS_\cG$ such that $\cC_l = \sigma \cdot \cC'_l$. To identify inequivalent cycles under $\cS_\cG$ we use the following 
procedure:
\begin{itemize}
    \item Decompose the set of possible values of $V'_1$, 
    i.e.~$\{1_1, 2_1, \dots, z_1\}$, into distinct orbits under $\cS_\cG$:  
    \begin{equation}
        \{1_1, 2_1, \dots, z_1\} 
        = \bigsqcup_{\bar{V}'_1} \mathrm{Orb}_{\cS_\cG}(\bar{V}'_1),
    \end{equation}
    where $\bar{V}'_1$ are orbit representatives.  

    \item For each representative $\bar{V}'_1$, compute the stabilizer subgroup 
    $\mathrm{Stab}_{\cS_\cG}(\bar{V}'_1) \subset \cS_\cG$ that leaves 
    $\bar{V}'_1$ fixed.  

    \item Decompose the possible values of $V_2$ into orbits under 
    $\mathrm{Stab}_{\cS_\cG}(\bar{V}'_1)$, and proceed iteratively. More 
    generally, to fix $\bar{V}_k$, choose a representative of each orbit of 
    $\{1_k, 2_k, \dots, z_k\}$ under the stabilizer group 
    $\mathrm{Stab}_{\cS_\cG}(\bar{V}'_1, \bar{V}_2, \dots, \bar{V}'_{k-1})$.  
\end{itemize}

This iterative decomposition provides the complete set of inequivalent 
$l$-cycles, modulo color permutations and bubble automorphisms. This dramatically reduces the number of cycles to be studied.

\label{Section:Orbit decomposition and inequivalent cycles}

\section{Intermediate field method for sextic interactions}
\label{Sec3}


As already mentioned in the introduction, the intermediate field method extends to tensorial interactions. 
This was already implemented for quartic interactions in 
\cite{lionni2016intermediate}.
Moreover, this method was also 
successfully applied to the prismatic interaction in 
\cite{Giombi_2018, Krajewski_2023}.
In this section we 
generalize the implementation of this method to the connected sextic 
interactions, except the wheel interaction. For this purpose, we use either a real intermediate tensor field or a complex intermediate tensor field.   


\subsection{Real intermediate field}

\begin{figure}[H]
    \centering
    \includegraphics[width=0.2\textwidth]{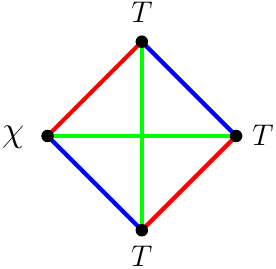}
    \caption{Tetrahedric term $\hat{I}_T(T,\chi)$ with an intermediate field.}
    \label{fig:IT_CHI}
\end{figure}
\begin{figure}[H]
    \centering
    \includegraphics[scale=0.7]{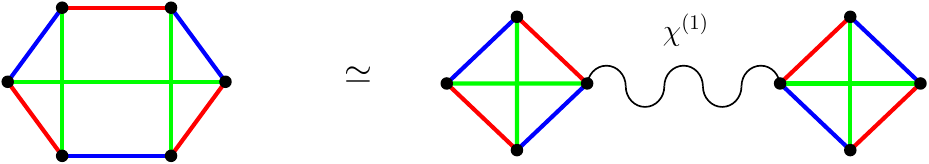}
    \caption{Decomposition of the prismatic interaction using a real intermediate field $\chi^{(1)}_{abc}$.}
    \label{fig:decomposition prismatic}
\end{figure}

One rewrites the $I_1$ interaction in the following way:

\begin{equation}
    \exp\Big[-\frac{g_1N^{3/2}}{6}I_1(T)\Big]=    \int[\text{d}\chi^{(1)}]\exp\Big[-N^{3/2}\big(\frac{1}{2}\chi_{ijk}^{(1)}\chi_{ijk}^{(1)}+i\sqrt{\frac{g_1}{3}}\hat{I}_T(T,\chi)\big)\Big],
\end{equation}
where 
\begin{equation}
    \hat{I}_T(T,\chi^{(1)})=T_{a_1b_1c_1}T_{a_1b_2c_2}T_{a_2b_1c_2}\chi_{a_2b_2c_1}^{(1)}
    \label{I_TCHI}
\end{equation}
is a tetrahedric bubble represented on Fig. \ref{fig:IT_CHI}, and $\chi^{(1)}$ is the intermediate field, here a rank $3$ tensor field. 
Therefore, a prismatic interaction can be decomposed in two tetrahedric interactions connected by an intermediate field $\chi_{abc}^{(1)}$, see Fig. \ref{fig:decomposition prismatic}.

\begin{figure}[H]
    \centering
    \includegraphics[scale=0.6]{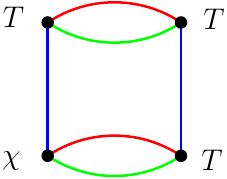}
    \caption{Pillow term $\hat{I}_P^{(a)}(T,\chi)$ with an intermediate field.}
    \label{fig:I_P-chi}
\end{figure}

In an analogous way, the interaction $I_4(T)$ can also be decomposed using a real intermediate field. The invariant is 
\begin{equation}
    I_4(T)=\frac{1}{3}(I_4^{(a)}(T)+I_4^{(b)}(T)+I_4^{(c)}(T)),
\end{equation}
where 
\begin{equation}
    I_4^{(a)}(T)=T_{ab_1c_1}T_{a_1bc}T_{a_1b_1c_1}T_{ab_2c_2}T_{a_2bc}T_{a_2b_2c_2},    
\end{equation}
 and the other terms are obtained by the color permutations $(a\Longleftrightarrow b)$ and $(a\Longleftrightarrow c)$.
Each of these bubble can be split as above, using the real intermediate field method.
\medskip 
For example, $I_4^{(a)}(T)$ can be expressed as $J_{abc}^{(a)}J_{abc}^{(a)}$ where $J_{abc}^{(a)}=T_{ab_1c_1}T_{a_1bc}T_{a_1b_1c_1}$. \\
The intermediate field is, in this case, $\chi_{abc}^{(4,a)}$.
The resulting term in the path integral is:


\begin{equation}
    \exp\Big[-\frac{g_4N^{3/2-1}}{6}I_4^{(a)}(T)\Big]=    \int[\text{d}\chi^{(4,a)}]\exp\Big[-N^{3/2}\big(\frac{1}{2}\chi_{ijk}^{(4,a)}\chi_{ijk}^{(4,a)}+\frac{i}{N^{1/2}}\sqrt{\frac{g_4}{3}}\hat{I}_P^{(a)}(T,\chi)\big)\Big],
\end{equation}
where $I_P^{(a)}(T,\chi)= T_{ab_1c_1}T_{a_1bc}T_{a_1b_1c_1}\chi^{(4,a)}_{abc}$ is a pillow bubble, represented on Fig. \ref{fig:I_P-chi}.  

\begin{figure}[H]
    \centering
    \includegraphics[scale=0.6]{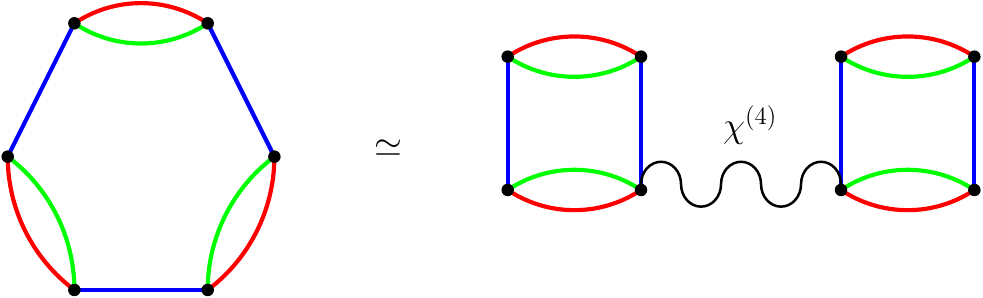}
    \caption{Decomposition of $I_4(T)$ by an intermediate field.}
    \label{fig:I_4 split}
\end{figure}

This gives the decomposition for the $I_4(T)$ interaction, as two pillows of the same color connected by a real intermediate field $\chi_{abc}^{(4,a)}$, see Fig. \ref{fig:I_4 split}.  

\subsection{Complex intermediate field}


The mechanism described in the previous subsection cannot be directly applied for the $I_3$ and $I_5$ interaction terms of the model studied here.
Nevertheless, for these interaction terms, we can adapt the previous intermediate field method using a complex intermediate field.


For these interactions, we define a complex intermediate field $\chi$ and its complex conjugate $\bar{\chi}$:
\begin{equation}
\chi_{abc}=\frac{1}{\sqrt{2}}(\psi_{abc}+i\phi_{abc}),\qquad \bar{\chi}_{abc}=\frac{1}{\sqrt{2}}(\psi_{abc}-i\phi_{abc}).
\end{equation}

The $I_5(T)$ bubble is thus decomposed as such:

\begin{equation}
\begin{split}
      &\exp\Big[-\frac{g_5N^{\frac{3}{2}-\frac{1}{2}}}{6}I_5(T)\Big]=\\&\int[\text{d}\chi^{(5)}][\text{d}\bar{\chi}^{(5)}]\exp\Big[-N^{3/2}(\chi_{abc}^{(5)}\bar{\chi}_{abc}^{(5)}+i\sqrt{\frac{g_5}{6}}\hat{I}_T(T,\chi)+iN^{-1/2}\sqrt{\frac{g_5}{6}}\hat{I}_P(T,\bar{\chi}))\Big],
\end{split}
\end{equation}
where $\hat{I}_T(T,\chi)$ is defined in eq. (\ref{I_TCHI}) and $\hat{I}_P(T,\bar{\chi})$ is a pillow bubble.

\begin{figure}[H]
    \centering
    \includegraphics[scale=1.2]{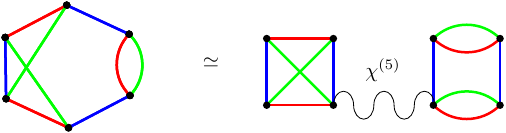}
    \caption{Intermediate field method for the  $I_5$ sextic interaction.}
    \label{fig:decompo g5}
\end{figure}

Thus, the $I_5$ interaction is split into a tetrahedron and a pillow interaction. This splitting is shown on Fig. \ref{fig:decompo g5}.

For the $I_3(T)$ interaction, a normalization over color permutation is again necessary. One has 
\begin{equation}
    I_3(T)=\frac{1}{3}(I_3^{(a)}(T)+I_3^{(b)}(T)+I_3^{(c)}(T)),
\end{equation}
where 
\begin{equation}
    I_3^{(a)}(T)=T_{a_1b_1c_1}T_{a_1b_1c_2}T_{a_2b_2c_2}T_{a_3b_2c_3}T_{a_3b_3c_3}T_{a_2b_3c_1},
\end{equation}
 and the other terms are again obtained by the color permutations $(a\Longleftrightarrow b)$ and $(a\Longleftrightarrow c)$.

Analogously, the $I_3^{(a)}$ bubble can be split as

\begin{equation}
\begin{split}
         &\exp\Big[-\frac{g_3N^{3/2-1}}{6}I_3^{(a)}(T)\Big]=\\
         &\int[\text{d}\chi^{(3,a)}][\text{d}\bar{\chi}^{(3,a)}]\exp\Big[-N^{3/2}(\chi_{abc}^{(3,a)}\bar{\chi}_{abc}^{(3,a)}+iN^{-1/2}\sqrt{\frac{g_3}{6}}\hat{I}_P^{(b)}(T,\chi)+iN^{-1/2}\sqrt{\frac{g_3}{6}}\hat{I}_P^{(c)}(T,\bar{\chi}))\Big].
\end{split}
\end{equation}



\begin{figure}[H]
    \centering
    \includegraphics[scale=0.6]{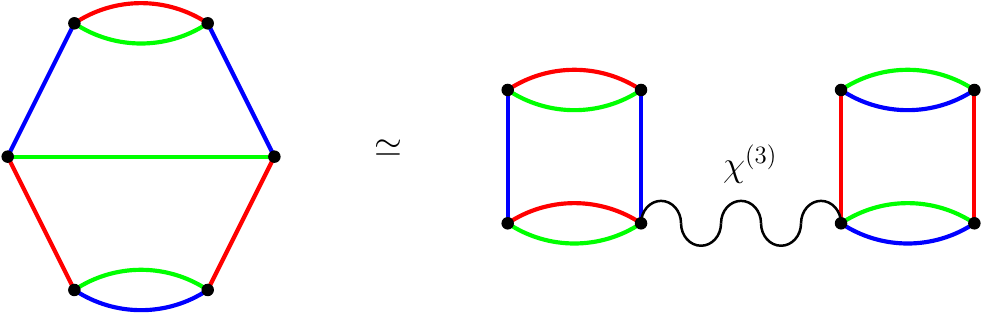}
    \caption{Intermediate field method for the  $I_3$ sextic interaction.}
    \label{fig:IFM-split-g3}
\end{figure}

This provides a splitting of $I_3(T)$ in two pillow interactions of different colors (see Fig. \ref{fig:IFM-split-g3}). 
Note that in Fig. \ref{fig:IFM-split-g3} a sum over color permutations is again implied.\\


Thus, the method we described here allows to reduce the order of the interactions in the model, from sextic to quartic, for all the interactions except $I_2$ (not that interactions $I_6, I_7, I_8$, can be 
considered 
by definition 
of lower order, since the respective bubbles  are not connected).
This is a powerful tool to identify the behavior of the sextic bubbles in the large $N$ limit since, the sextic large $N$ limit can now be reconstructed from the known behavior of the quartic large $N$ limit.


Note that the analysis above allows us to rewrite our general sextic model as a model with the wheel sextic interactions and a set of quartic interactions (the tetrahedron, the pillow) and the trace bubble, but with a non-trivial set of intermediate real or complex tensor fields ($\chi^{(1)}, \chi^{(3)}, \chi^{(4)}$ and $\chi^{(5)}$). We refer to the model with the sextic interactions and only one type of propagator as the \textit{sextic representation} and the model with the quartic interactions and many propagators as the \textit{quartic representation}.


\section{No mixing of the wheel interaction}
\label{Sec4}
The first question one needs to address is whether or not one can find dominant fundamental graphs of the model which contain both the wheel interaction and quartic interactions. We prove in this section that no such mixing of interactions is allowed if the respective graph is dominant.


This holds for dominant fundamental graph $\cG$. As usual in the tensor literature, we call a fundamental dominant graph a graph $\omega(\cG)= 0$ from which all the dominant graphs can be constructed \textit{via} $2$-points or $6$-points insertions.

\begin{figure}[H]
    \centering
    \includegraphics[width=0.6\linewidth]{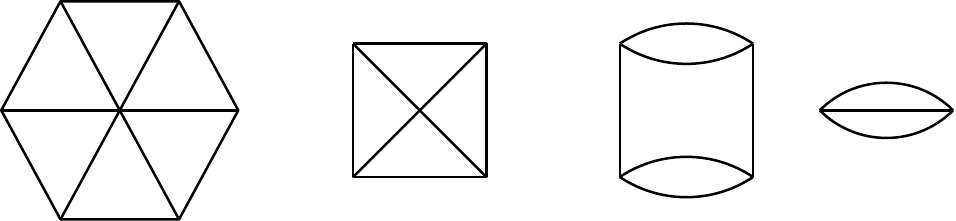}
    \caption{The reduced set of interactions that builds the sextic invariants.}
    \label{fig:reduced_set}
\end{figure}

Note that, as a consequence of the analysis of the previous section, our model can now be studied as a model with a reduced set of interactions, namely the wheel, the tetrahedron, the pillow and the trace (Fig. \ref{fig:reduced_set}). 

In the case of $0-$dimensional field theory, the computations of Feynman amplitudes is a pure combinatorial computation. Thus, we prove that the model with only one type of propagator instead of $5$ types of propagators 
($T, \chi^{(1)}, \chi^{(3)}, \chi^{(4)}$ and $\chi^{(5)}$)
does not allow mixing between the wheel interaction and the tetrahedron interaction. This implies that the original model with $5$ types of propagators does not allow this mixing either.

 In order to simplify the notation, we thus choose to have a single type of field in our model. Later on, in order to identify explicitly the dominant graphs, we will come back to the original $5$ field model. 


\medskip

\begin{figure}[H]
    \centering
    \includegraphics[width=0.6\linewidth]{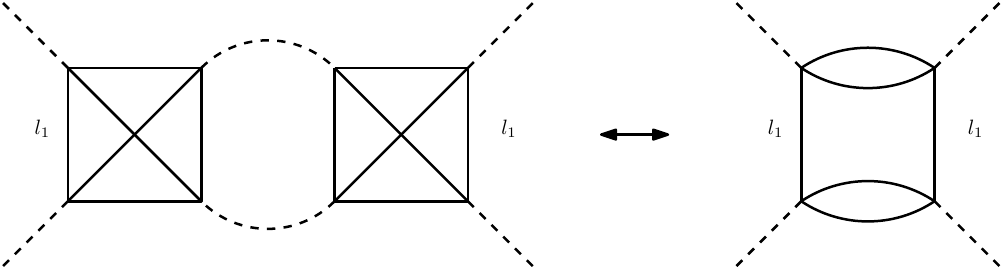}
    \caption{A replacement of a pillow by two tetrahedron.}
    \label{fig:pillow=2tetra}
\end{figure}

\begin{figure}[H]
     \centering
     \includegraphics[width=0.6\linewidth]{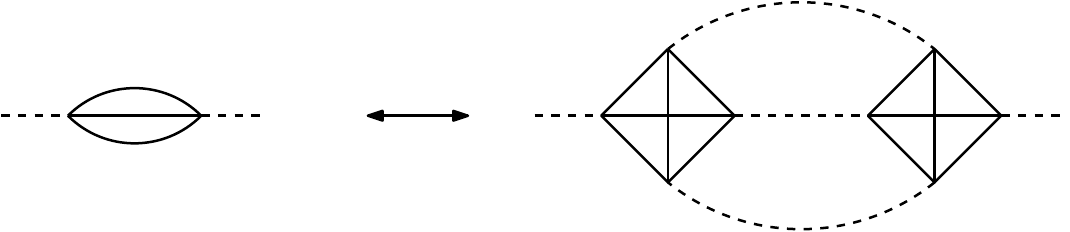}
     \caption{A replacement of a trace by two tetrahedron.} 
     \label{fig:trace=2tetra}
 \end{figure}

The first step of the proof of the no mixing result is to
further reduce the set of interactions by identifying two particular moves that preserve the degree. 
As already noticed in \cite{Carrozza_2016}, the pillow 
interaction 
can be represented by two tetrahedric bubbles. Indeed, one can directly prove that the move 
defined in Fig. \ref{fig:pillow=2tetra}
doesn't change the degree.
 The trace bubble also can be generated by tetrahedric bubbles, via the move
 defined in Fig. \ref{fig:trace=2tetra}.

The model can therefore be further reduced to an even simpler model with only wheel and tetrahedron interactions that are allowed. 
We now exhibit the dominant graphs of this reduced model.


We consider a dominant Feynman graph $\mathcal{G}$ containing both tetrahedric and wheel interaction, in the stranded representation. We first prove a bound on the number of faces of lengths $1$, $2$ and $3$.
We follow here \cite{Carrozza_2016} and consider the three Jackets $\mathcal{J}_i,~ i = 1,2,3$, obtained by removing all the edges of color $i$ from $\cG$. One can prove that the jackets constructed in this way are ribbon graphs. 
Counting the faces of each such a ribbon jacket yields:
\begin{equation}
    \sum_{i=1}^3F(\mathcal{J}_i)=2F(\mathcal{G}).
\end{equation}
The number of faces of each jacket is given by 
\begin{equation}
    F(\cJ_i)+V(\cG)-E(\cG)=\sum_{n=1}^{N_i} 2-2h(\cJ_i^{(n)}),
\end{equation}
where $N_i$ is the number of connected component of $\cJ_i$, $h(\cJ_i^{(n)})$ is the non orientable genus of the $n$-th component of the $i$-th jacket and $V(\cG)$ and $E(\cG)$ denote the vertices, respectively the edges of the tensor Feynman graph $\cG$. The tensor Feynman graph $\cG$ is dominant if all its ribbon jackets have genus $0$ (and are thus planar). \\
We only consider tetrahedric and wheelic interactions such that
\begin{equation}
        V(\cG)=v_t+v_w ~~\text{and} ~~ E(\cG)=\frac{1}{2}(4v_t+6v_w)=2v_t+3v_w,
\end{equation}
where $v_t$ and resp. $v_w$ are the numbers of tetrahedric and resp. wheelic bubbles. 
We can thus express the total number of faces as 
\begin{equation}
    F(\cG)=\frac{3}{2}v_t+3v_w+(N_1+N_2+N_3).
\end{equation}
Note that if $\cG$ is a dominant graph, then it contains an even number of tetrahedric bubbles, as in the $O(N)^3$-invariant quartic model.

The total number of faces can also be expressed as a sum of faces of length $l$: 
\begin{equation}
    F(\cG)=\sum_{l\geq 1} F_l.
    \label{fl}
\end{equation}
One then has
\begin{equation}
   \sum_{l\geq} lF_l=3E(\cG)=6v_t+9v_w,
   \label{lfl}
\end{equation}
which further leads to
\begin{equation}
    \sum_{l\geq 1}(4-l)F_l=4(N_1+N_2+N_3)+3v_w.
\end{equation}
This further rewrites as:
\begin{equation}
\begin{split}
        3F_1+2F_2+F_3&=4(N_1+N_2+N_3)+3v_w+ \sum_{l\geq 5}(l-4)F_l.\\
     \end{split}    
\label{boundFaces}
\end{equation}
Since each term on the RHS of the eq. (\ref{boundFaces}) above is positive, one has the following bound:
\begin{equation}
\begin{split}
        3F_1+2F_2+F_3 \ge 12 \ge 1
     \end{split}    
\label{boundFaces2}
\end{equation}

Therefore a dominant graph 
has at least one face visiting one, two or three bubbles. Following the lines of \cite{Prakash_2020}, we use a weaker version 
of this bound, namely we use the fact that a  dominant graph contains at least a cycle of length one, two or three. 
This comes from the fact a cycle is not necessarily a face in the bubble representation (see again Fig. \ref{fig:cycle_exemple}).

Note that dominant graphs containing cycles visiting either only tetrahedric or only wheel bubbles
have already been studied in \cite{Carrozza_2016, Prakash_2020}. 

Let us first recall the results for these dominant graphs containing cycles visiting either only tetrahedric or only wheel bubble.

\begin{figure}[H]
    \centering
    \includegraphics[scale=0.6]{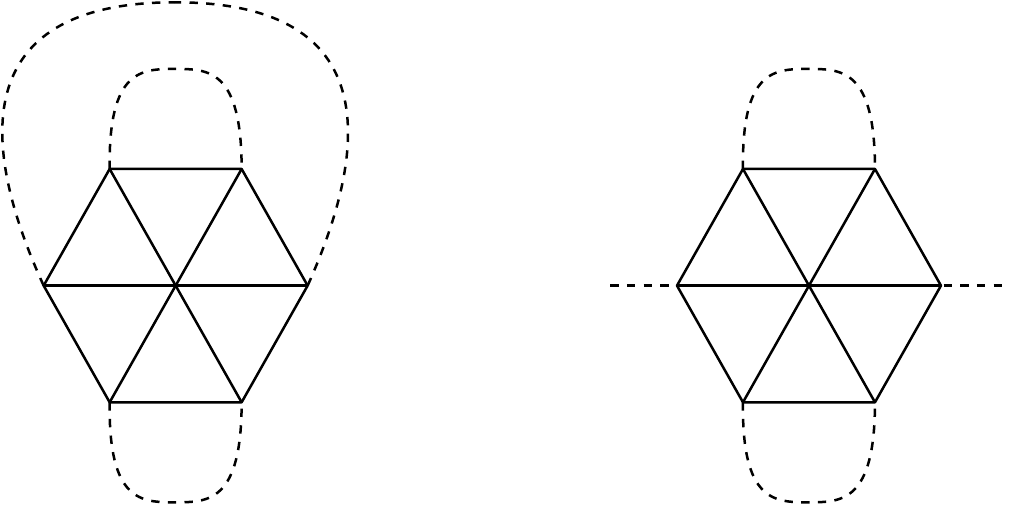}
    \caption{Fundamental dominant graphs containing a 1-cycle visiting a wheel bubble (left) and a corresponding tadpole insertion (right).}
    \label{fig:cycle1wheel}
\end{figure}

 No dominant graph contains a cycle of length one visiting a tetrahedron (see again \cite{Carrozza_2016}).
The only fundamental dominant graph containing a cycle visiting one wheel is the triple tadpole. Its form is given on Fig. \ref{fig:cycle1wheel} (see again
\cite{Prakash_2020}).

\begin{figure}[H]
    \centering
    \includegraphics[width=0.7\linewidth]{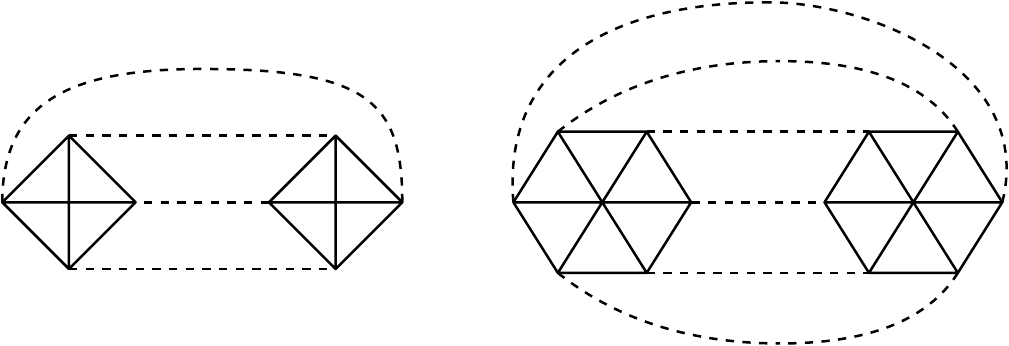}
    \caption{Fundamental dominant graphs containing a 2-cycle $TT$ (left) and $WW$ (right).}
        \label{fig:2cyclesMelon}
\end{figure}

The 2-cycles can visit two tetrahedron ($TT$), two wheels ($WW$) or one tetrahedron and one wheel bubble ($WT$). The dominant graphs containing a $TT$ cycle or a $WW$ cycle have again been found in \cite{Carrozza_2016, Prakash_2020}. These are the melonic  graphs (Fig. \ref{fig:2cyclesMelon}). 

\begin{figure}[H]
    \centering
    \includegraphics[width=0.7\linewidth]{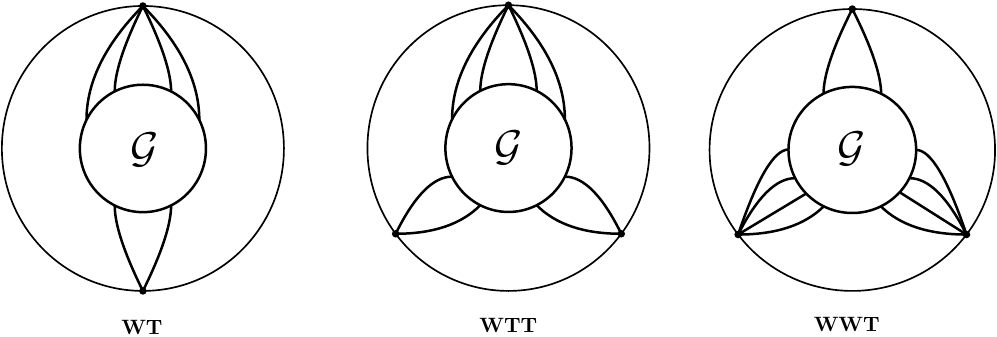}
    \caption{General shape of dominant graphs containing $WT$, $WTT$ or $WWT$ cycles.}
    \label{fig:mixedCycles}
\end{figure}

A dominant graph might also contain a 2-cycle visiting a tetrahedron and a wheel (a $WT$ cycle). The general form of such graph is shown of Fig. \ref{fig:mixedCycles} (left). 

The $TTT$ and resp. the $WWW$ cycles have again been studied in \cite{Carrozza_2016} and resp. \cite{Prakash_2020}.
No fundamental dominant graph contains a $TTT$ cycle. 
Analogously, no fundamental dominant graph contains a $WWW$ cycle.
However, a dominant graph containing a $WWW$ cycle can only be constructed using the  insertion of Fig.~\ref{fig:cycle1wheel} (right)
on a propagator of the graph of Fig. \ref{fig:2cyclesMelon} (right). 
Note that this type of insertion cannot be done for the tetrahedric melon on the left of Fig. \ref{fig:2cyclesMelon}
since one needs to insert in this case a $2-$point function containing (at least) two tetrahedric bubbles.

\bigskip

Let us now analyze the cases of  mixed interaction:
\begin{itemize}
    \item $2$-cycles with  $WT$
\item $3$-cycles visiting mixed interactions ($WTT$ and $WWT$) can appear in dominant graphs.
\end{itemize}

 We sum up all the possible cycles visiting mixed types of interactions that might appear in a dominant graph on Fig. \ref{fig:mixedCycles}.

\begin{figure}[H]
    \centering
    \includegraphics[scale=0.7]{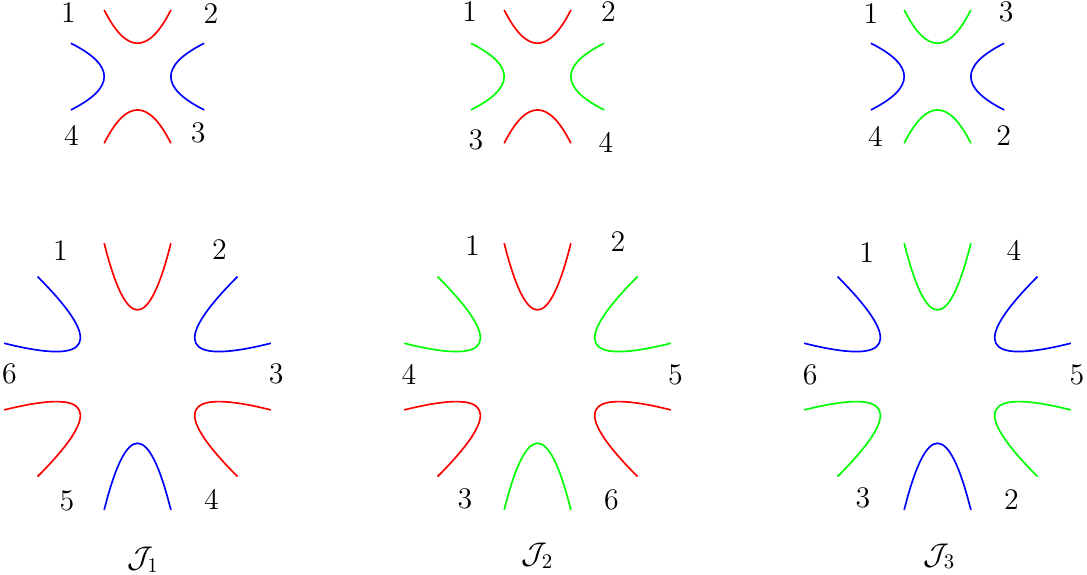}
    \caption{Jackets for the tetrahedron and the wheel bubble. The external bubble vertices are labeled.} 
    \label{fig:WTRibbon}
\end{figure}

To identify the dominant graphs of the form given in Fig. \ref{fig:mixedCycles}, we follow a procedure introduced in \cite{Prakash_2020}, where this is used to identify the large $N$ limit of a sextic model containing only prismatic and wheel interactions. 
The procedure is the following:
\begin{itemize}
    \item Take one of the graphs in Fig. \ref{fig:mixedCycles} and list all its possible inequivalent cycles (see subsection \ref{Section:Orbit decomposition and inequivalent cycles}).
    \item For each inequivalent cycle, impose planarity for each ribbon jacket $J_l$ (see Fig.\ref{fig:WTRibbon}). In particular, remove the cycles that contain an odd number of twists. 
    \item Planarity of each jacket imposes constraints on $\cG$ from Fig.\ref{fig:mixedCycles}. Only dominant graphs satisfy these constraints.  
\end{itemize}

\paragraph{The WT cycle.}

\begin{figure}[H]
    \centering
    \includegraphics[scale=0.6]{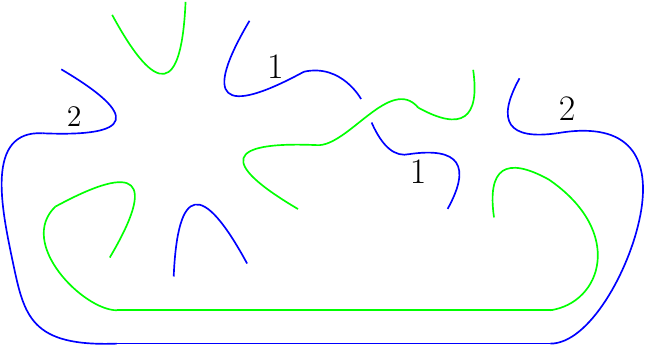}
    \caption{The $\cJ_3$ jacket of a graph containing the $\langle(1_L,1_R),(2_R,2_L)\rangle$ cycle.}
    \label{fig:WTjac}
\end{figure}

We find two inequivalent cycles with respect to color permutation cycles and bubble automorphisms:
\begin{itemize}
    \item $\langle(1_W,1_T),(2_W,2_T)\rangle$
    \item $\langle(1_W,1_T),(2_W,4_T)\rangle$
\end{itemize}
These two cycles are not dominant because some of their ribbon jackets have an odd number of twists. 
For example, when we draw the first cycle in the $\cJ_3$ jackets, we find that it has one twist (see Fig. \ref{fig:WTjac}). Therefore the corresponding graph containing this cycle in the stranded representation cannot be dominant. 

This is also the case for the second cycle. We thus find that no dominant  graph contains a $WT$ cycle. 

\paragraph{The WTT cycle.}

\begin{figure}[H]
    \centering
    \includegraphics[width=0.4\textwidth]{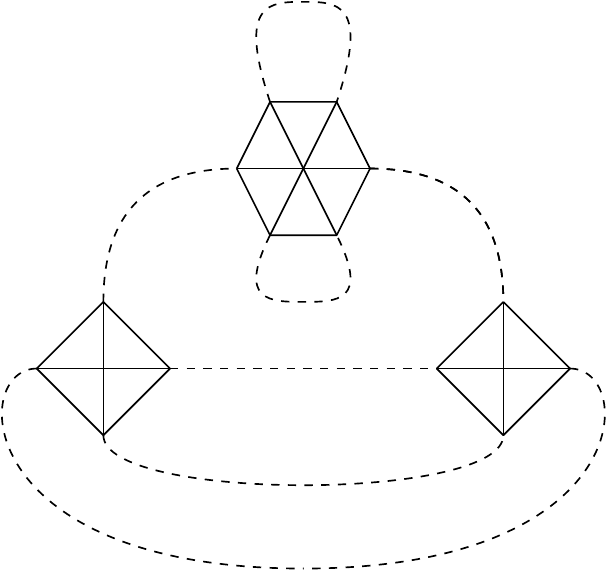}
    \caption{Dominant graph containing a $WTT$ cycle.}
    \label{fig:WTTok1}
\end{figure}

We find five inequivalent cycles visiting two wheels and a tetrahedron:
\begin{itemize}
    \item $\langle(1_W,1_T),(2_T,1_{T'}),(\chi_{T'},2_W)\rangle$
    \item $\langle(1_W,1_T),(4_T,1_{T'}),(\zeta_{T'},2_W)\rangle$
\end{itemize}
where $\chi=2, 3, 4$ and $\zeta=2, 4$. 
One has two allowed cycles: $\langle(1_W,1_T),(2_T,1_{T'}),(2_{T'},2_W)\rangle$ and  $\langle(1_W,1_T),(4_T,1_{T'}),(4_{T'},2_W)\rangle$. 
Planarity of each jacket constrains again the graphs containing these allowed cycles to have the generic shape shown on Fig. \ref{fig:WTTok1}. This graph corresponds to a melonic graph built on two tetrahedron with a $2$-point wheelic insertion on a propagator (or equivalently to a triple tadpole with a $2$-point melonic insertion). The wheel insertion can be performed on every propagator of the melonic graph.

Note that the allowed $WTT$ cycles do not generate new fundamental structures, since graphs of the form given in Fig. \ref{fig:WTTok1} can be obtained from other fundamental graphs and insertions.

\paragraph{The WWT cycle.}

Graphs containing $WWT$ cycles have the generic form given Fig. \ref{fig:mixedCycles} (right). 

One can prove that $17$ inequivalent cycles can be built using two wheel and a tetrahedron bubble:

\begin{itemize}
    \item $\langle(1_W,1_{W'}),(2_{W'},1_T),(2_T,\alpha_W)\rangle$
    \item $\langle(1_W,1_{W'}),(2_{W'},1_T),(3_T,\beta_W)\rangle$
    \item $\langle(1_W,1_{W'}),(3_{W'},1_T),(\beta_T,\gamma_W)\rangle$
\end{itemize}
with $\alpha \in \{2, 3, 4, 5, 6\}$, $\beta \in \{2, 3, 4\}$ and $\gamma \in \{2, 3, 5\}$.

All these cycles have an odd number of twists in at least one of their ribbon jackets. This means that no dominant graph can contain a $WWT$ cycle.

\bigskip

Let us now sum up the results obtained so far. We argued that every sextic bubble (except the wheel) 
can be decomposed in tetrahedric bubbles 
(using either the intermediate field method or the melonic moves of Fig. \ref{fig:pillow=2tetra} and \ref{fig:trace=2tetra}). 
We further proved that dominant graphs from this reduced model, built out of tetrahedron and wheel interactions,
must contain at least one cycle of length $1$, $2$ or 
$3$. 

We find that the dominant graphs in this reduced model are the triple-tadpole graphs given in  Fig. \ref{fig:cycle1wheel}, the melon graphs given in Fig. \ref{fig:2cyclesMelon} and the graph given in Fig. \ref{fig:WTTok1}. However, the latter is not a fundamental dominant graph since it can be obtained from a $2-$point wheel insertion (which conserves the degree) on a fundamental  melon. 
This allows us to conclude that there are no fundamental dominant graphs containing both the wheel and the tetrahedron interaction. As a consequence, we have thus proved in this section, that there are no fundamental dominant graph that contains both the wheel and another sextic bubble.  

\section{Identification of the dominant graphs of the general model}
\label{Sec5}

For each bubble $\cB$, we list the fundamental dominant graphs and the allowed insertions that 
conserve the degree.
We then explicitly identify dominant graphs where several types of sextic bubbles, except the wheel one, can appear within the same graph.


\subsection{The prismatic interaction}

\begin{figure}[H]
    \centering
    \includegraphics[scale=0.6]{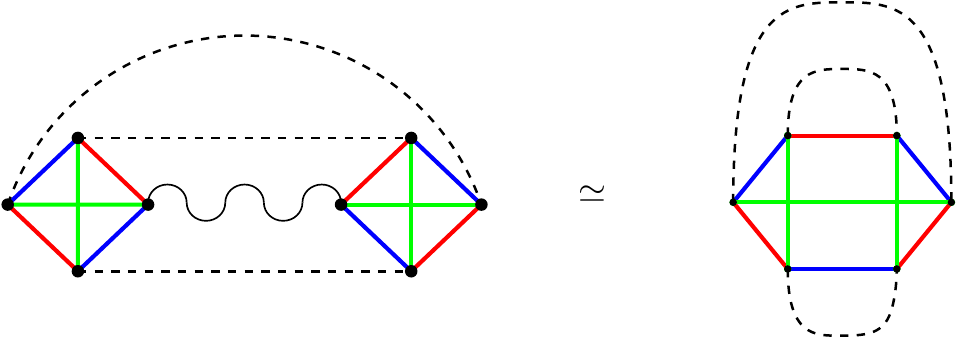}
    \caption{Dominant graph in the tetrahedric (left) and prismatic representation (right).}
    \label{fig:fundamental tetra}
\end{figure}

The large $N$ limit of the prismatic bubble was studied in \cite{Giombi_2018} with an intermediate field method. 
The only dominant fundamental graph is the graph of  Fig. \ref{fig:fundamental tetra}, where both the  
tetrahedric and prismatic representations are given. Note that, in the prismatic representation this fundamental dominant graph is a triple tadpole.

\begin{figure}[H]
    \centering
    \includegraphics[scale=0.5]{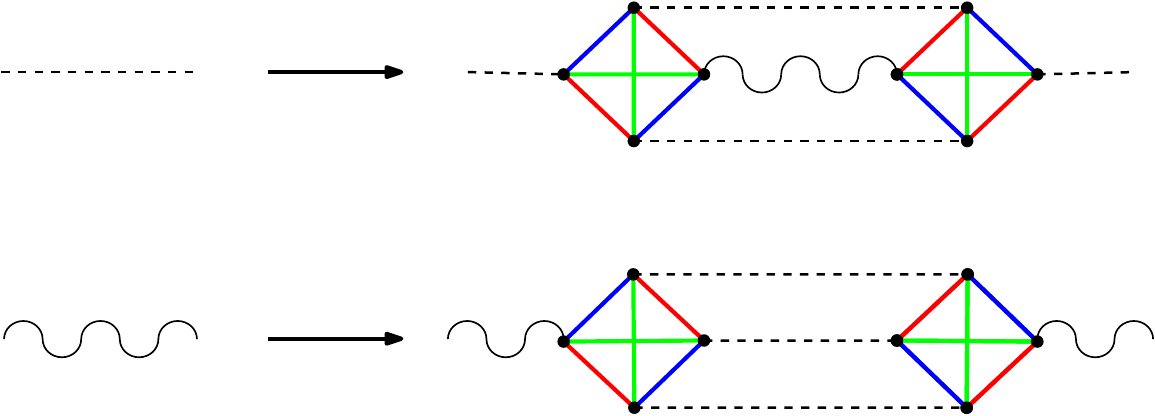}
    \caption{Melonic insertions on the $T_{abc}$ and the $\chi^{(1)}_{abc}$ propagator.}
    \label{fig:tetrahedric insertions}
\end{figure}

Following \cite{Krajewski_2023},
 and using the quartic representation, one can perform two kinds of insertions which conserve the degree of the graph: 
\begin{enumerate}
\item a melonic insertion on the $T_{abc}$ propagator and 
\item a melonic insertion on the  $\chi^{(1)}_{abc}$ propagator (see Fig. \ref{fig:tetrahedric insertions}).
\end{enumerate}

\begin{figure}[H]
    \centering
    \includegraphics[scale=0.7]{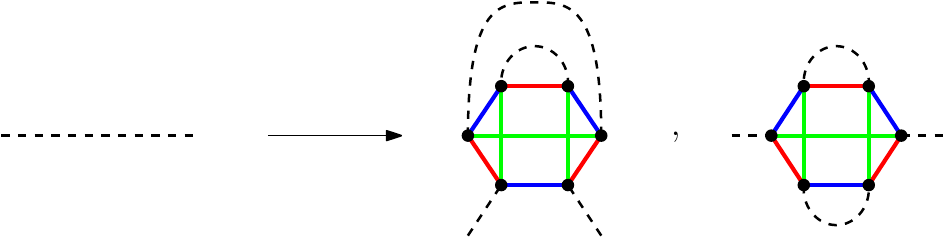}
    \caption{Double tadpole insertion on the $T_{abc}$ propagator.}
    \label{fig:g1 double tadpole insertion}
\end{figure}

\begin{figure}[H]
    \centering
    \includegraphics[scale=0.55]{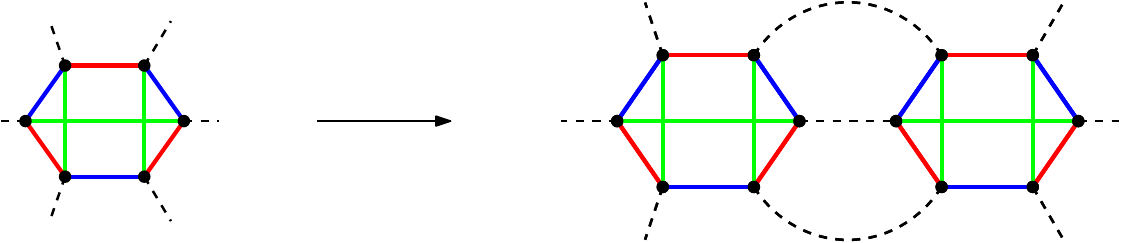}
    \caption{Melonic insertion in the $I_1$ bubble in the prismatic representation.}
    \label{fig:vertex insertion}
\end{figure}

Using now the prismatic representations, two types of insertions are allowed:
\begin{enumerate}
    \item The melonic insertions on $T_{abc}$ propagator of Fig.~\ref{fig:tetrahedric insertions} translates as a double tadpole insertion shown in Fig.~\ref{fig:g1 double tadpole insertion}. Note that the two double tadpoles of  Fig.~\ref{fig:g1 double tadpole insertion} are not equivalent from the point of view of index contractions.
    \item The melonic insertion on the $\chi^{(1)}_{abc}$ propagator of Fig.~\ref{fig:tetrahedric insertions} translates as an insertion at the level of the  $I_1$ bubble
    (see Fig.~\ref{fig:vertex insertion}).
\end{enumerate}

Note that the insertion in Fig.~\ref{fig:vertex insertion} is a vertex insertion, as already explained in \cite{Krajewski_2023}.


 \subsection{The wheel interaction}

\begin{figure}[H]
     \centering
     \includegraphics[scale=0.8]{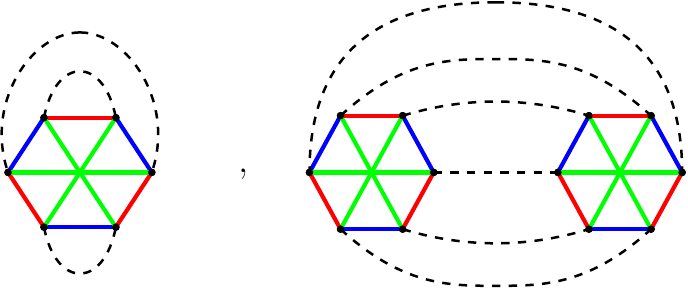}
     \caption{Fundamental dominant graphs built on $g_2$. The triple tadpole (left) and the melon (right).}
     \label{fig:FundaVacuumWheel}
 \end{figure}

 The fundamental dominant graphs containing only wheel interactions  have been found in section \ref{Sec4}. We list them in Fig. \ref{fig:FundaVacuumWheel}. 
 
\begin{figure}[H]
     \centering
     \includegraphics[scale=0.8]{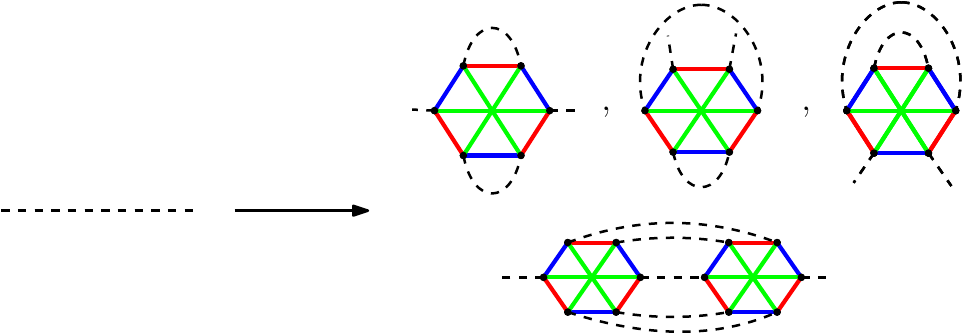}
     \caption{$2$-point insertion for the $I_2$ interaction.}
     \label{fig:FundaWheel2p}
 \end{figure}

 The $2-$point insertions conserving the degree are shown on Fig. \ref{fig:FundaWheel2p}. These insertions are obtained by cutting an edge in the fundamental dominant graphs. 
 
 \subsection{\texorpdfstring{The $I_3$ and $I_4$ interaction}{The I3 and I4 interaction}}

\begin{figure}[H]
     \centering
     \includegraphics[width=0.4\textwidth]{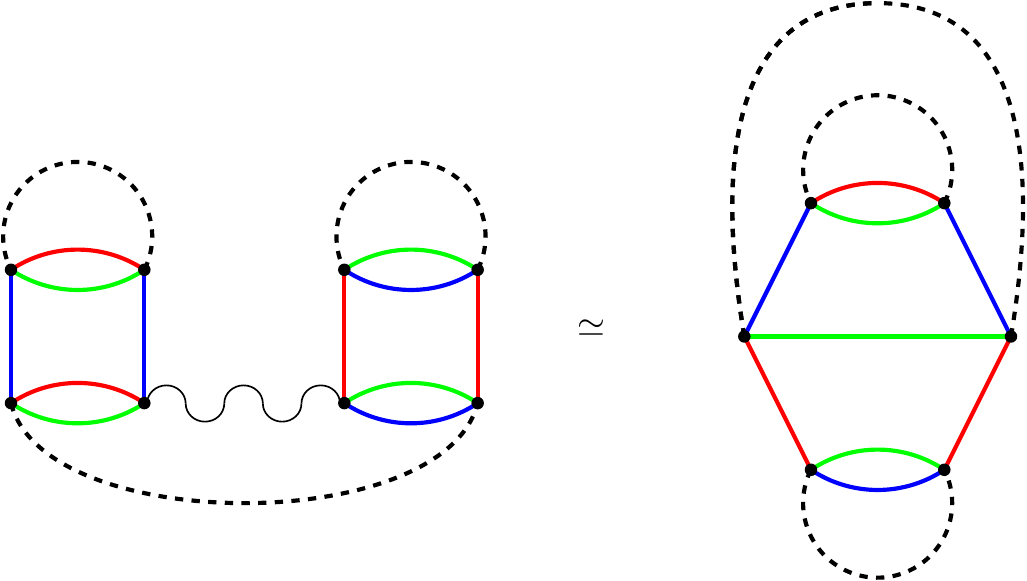}
     \caption{Fundamental graph from $I_3$ in the quartic (left) and sextic representation (right).}
     \label{fig:fundaVacg3}
 \end{figure}

\begin{figure}[H]
    \centering
    \includegraphics[width=0.6\textwidth]{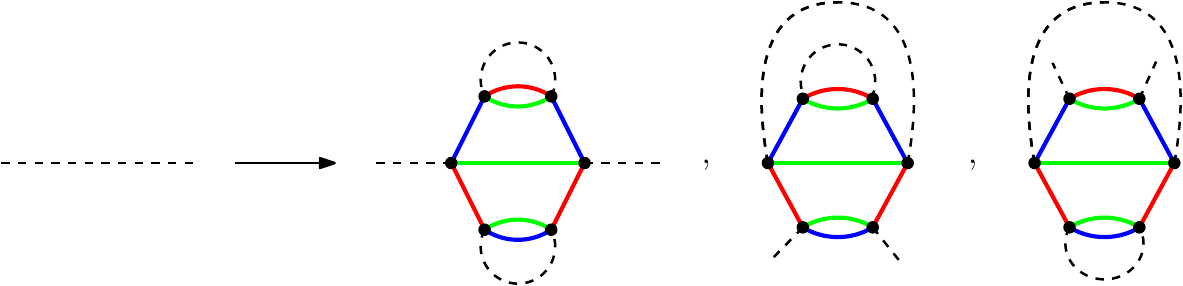}
    \caption{$2-$points insertions for the $I_3$ interaction.}
    \label{fig:funda2pg3}
\end{figure}

 Recall that the $I_3$ interaction can be split, using the intermediate field method, in two pillow interactions of different colors (see Fig. \ref{fig:IFM-split-g3} above). 
 The only dominant graph one can construct using only this interaction is
 given, in the quartic and sextic representation, in Fig. \ref{fig:fundaVacg3}.
Note that, once again, the dominant graph in the 
sextic representation is a triple tadpole.
 The insertions 
 conserving the degree
 are, once again, obtained by cutting one edge in the dominant graphs. 
 This can be done in three distinct ways (Fig. \ref{fig:funda2pg3}).
 
\begin{figure}[H]
    \centering
    \includegraphics[width=0.5\textwidth]{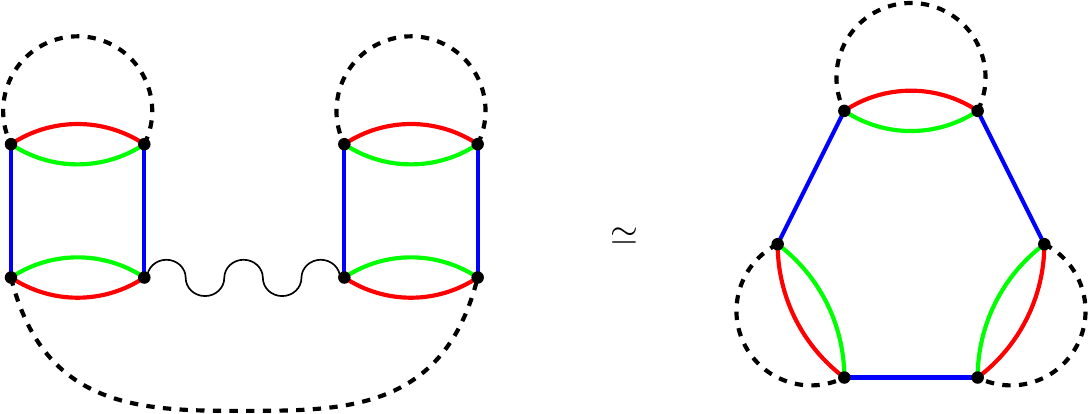}
    \caption{Dominant graph from $I_4$ in the quartic (left) and sextic representation (right).}
    \label{fig:fundaVacg4}
\end{figure}

As above, recall that 
the $I_4(T)$ interaction is split, using the intermediate field method, in two pillow interactions which are of the same color (see FIg. \ref{fig:I_4 split}). 
The only dominant graph one can construct using only this interaction is, once again, the triple tadpole.
We give this dominant graph in the quartic and sextic representation, in Fig. \ref{fig:fundaVacg4}.

\begin{figure}[H]
     \centering
     \includegraphics[width=0.5\textwidth]{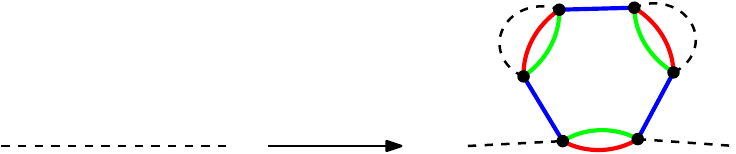}
     \caption{$2-$point insertions for the $I_4(T)$ interaction.}
     \label{fig:funda2pg4}
 \end{figure}

 The insertions conserving the degree is obtained by cutting an edge in the dominant graph. Only one insertion is allowed (Fig. \ref{fig:funda2pg4}).
 
 \subsection{\texorpdfstring{The $I_5$ and $I_6$ interaction}{The I5 and I6 interaction}}
\begin{figure}[H]
    \centering
    \includegraphics[width=\linewidth]{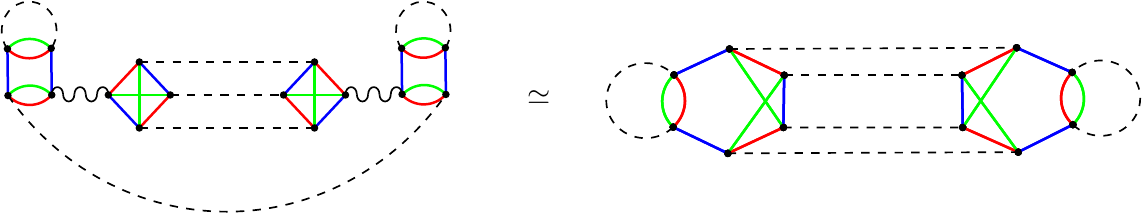}
    \caption{Dominant graph formed with $I_5(T)$.}
    \label{fig:g5_vac}
\end{figure}

 As above, recall that 
the $I_5(T)$ interaction is split, using the intermediate field method, in a pillow 
and  a tetrahedron
interactions - see \ref{fig:decompo g5}.
The only dominant graph one can now construct using only this interaction is
 given, in the quartic and sextic representation, in Fig. \ref{fig:g5_vac}.
 
 Let us emphasize here that the type of dominant graph of Fig. \ref{fig:g5_vac} is to our knowledge, new with respect to the types of dominant graphs known so far for various tensor models.
 
\begin{figure}[H]
     \centering
     \includegraphics[width=0.6\textwidth]{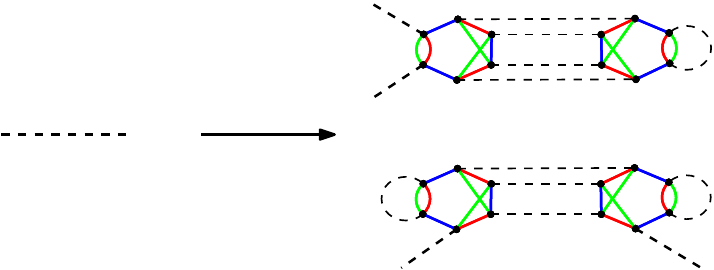}
     \caption{Dominant $2$-point insertions for the $I_5(T)$ interaction.}
     \label{fig:LOg5inser}
 \end{figure}

 The insertions conserving the degree are obtained again by cutting an edge in the dominant graph. Two insertions are allowed, see Fig. \ref{fig:LOg5inser}.
 
\begin{figure}[H]
    \centering
    \includegraphics{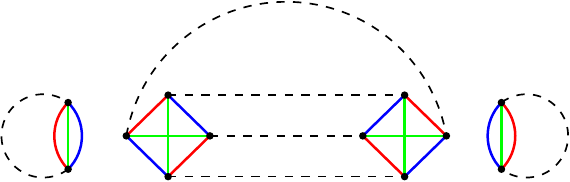}
    \caption{Dominant graph formed with $I_6(T)$.}
    \label{fig:g6_vac}
\end{figure}

Let us now analyze the 
$I_6$ interaction. Recall that this interaction is a disconnected one built on a tetrahedron and a trace. 
The dominant graphs are is thus the melon graph (formed with two $g_6$ bubbles) of 
Fig. \ref{fig:g6_vac}). 
  
\begin{figure}[H]
    \centering
    \includegraphics[width=0.6\textwidth]{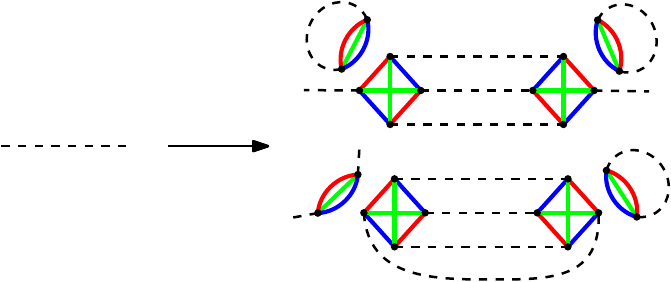}
    \caption{$2$-point insertions for the $I_6$ interaction.}
    \label{fig:LOg6@p}
\end{figure}

The $2-$point insertions conserving the degree are again obtained by cutting an edge in 
the dominant graph, see Fig. \ref{fig:LOg6@p}. 

 \subsection{\texorpdfstring{The $I_7$ and the $I_8$ bubbles}{The I7 and I8 bubbles}}
\begin{figure}[H]
    \centering
    \includegraphics[scale=0.5]{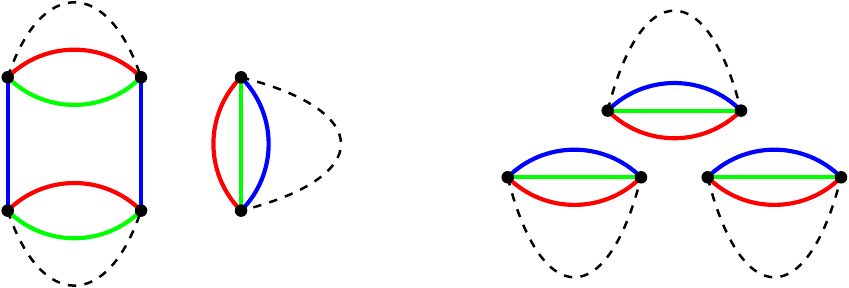}
    \caption{Dominant fundamental graphs for the $I_7(T)$ interaction (left) and the $I_8(T)$ interaction (right).}
    \label{fig:LOg78}
\end{figure}

The $I_7$ and $I_8$ bubbles are again disconnected and their large $N$ behavior 
is known. 
The fundamental dominant graphs are the triple tadpoles of Fig. \ref{fig:LOg78}. 

\begin{figure}[H]
    \centering
    \includegraphics[scale=0.5]{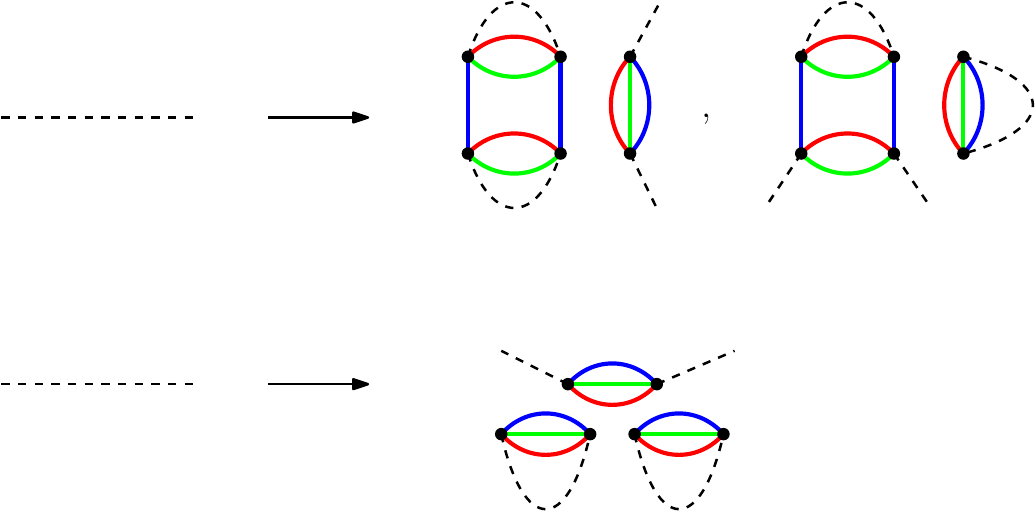}
    \caption{$2$-point insertions built $I_7(T)$ (top) and $I_8(T)$ (bottom).}
    \label{fig:LO2pg78}
\end{figure}

The  $2-$point insertions that conserve the degree are, once again, obtained by cutting an edge in any of the fundamental dominant graphs, see Fig. \ref{fig:LO2pg78}.

\subsection{Dominant graph containing several types of sextic interactions}

Note that, one can have several sextic interactions within the same fundamental dominant graph, except for the wheel one.
Let us emphasize that in the case of the 
$U(N)^3$ invariant model studied in \cite{Benedetti_2020} 
one cannot have dominant graphs containing several sextic interactions. However, this becomes possible for the $O(N)^3$ model studied here, as a consequence of the 
additional $I_1$, $I_5$ and $I_6$ interactions.

\begin{figure}[H]
    \centering
    \includegraphics[scale=0.8]{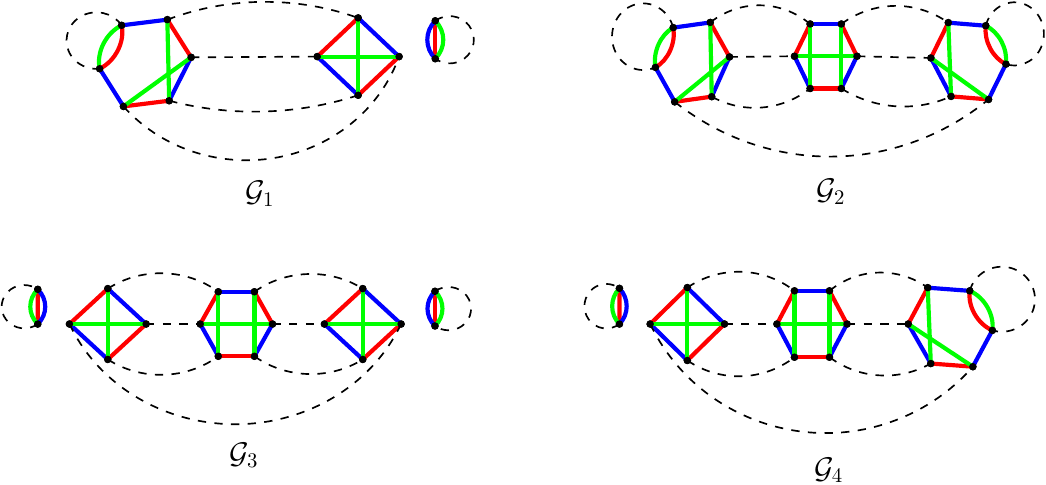}
    \caption{Fundamental dominant graphs with mixed interactions.}
    \label{fig:mixing fundamental}
\end{figure}

Using the intermediate field method and the large $N$ behavior for quartic interactions, we find four new fundamental dominant graphs containing mixed bubbles which we show on Fig. \ref{fig:mixing fundamental}. We denote these diagrams by $\mathcal{G}_1$ to $\mathcal{G}_4$.
Using eq. (\ref{omega6})
 we show that their degree is vanishing:
\begin{equation}
\begin{split}
    \omega(\mathcal{G}_1)&=3+\frac{7}{2}+\frac{9}{2}-11=0 ,\\
    \omega(\mathcal{G}_2)&=3+2\times\frac{7}{2}+3-13=0 , \\
    \omega(\mathcal{G}_3)&=3+2\times\frac{9}{2}+3-15=0 ,\\
    \omega(\mathcal{G}_1)&=3+\frac{7}{2}+\frac{9}{2}+3-14=0 .\\
\end{split}
\end{equation}

  Let us note that, one can use any of the new types of dominant graph of Fig. \ref{fig:mixing fundamental} to generate a series of non-fundamental dominant graphs, which are obtained, as usual, \textit{via}
 $2$-point insertions which preserve the degree. 
The pattern to be inserted is obtained, as usually, by cutting an edge in any of these fundamental vacuum graphs.


\subsection{Summary of the large $N$ limit results}

\begin{figure}[H]
    \centering
    \includegraphics[scale=0.8]{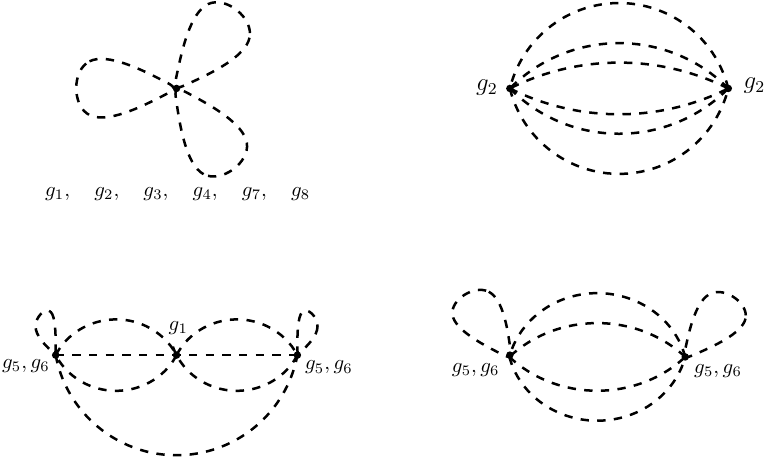}
    \caption{Dominant fundamental graphs for the $O(N)^3$-invariant sextic model.}
    \label{vacuumLO}
\end{figure}

We list here all the possible fundamental dominant graphs of our general sextic model (\ref{action}). This list of Feynman graphs is given in Fig. 
\ref{vacuumLO}, where we indicate which interactions are allowed in each case. 

\begin{figure}[H]
    \centering
    \includegraphics[scale=0.7]{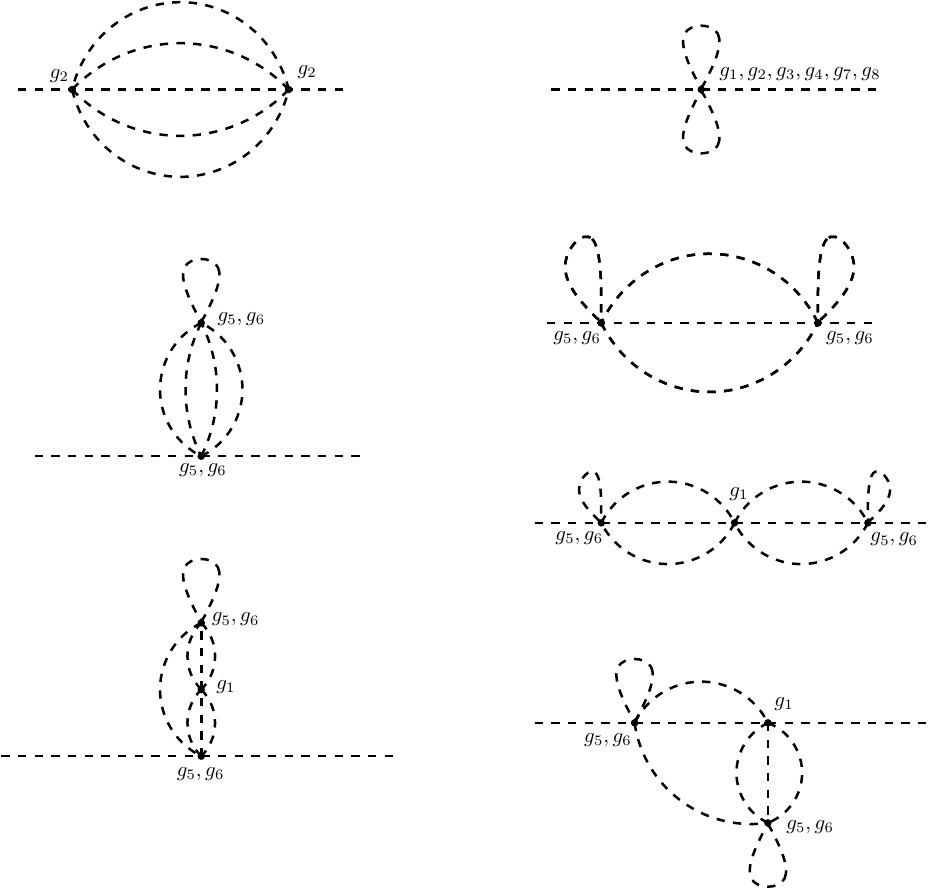}
    \caption{Possible degree-preserving insertions on the propagator.}
    \label{fig:covariance insertions}
\end{figure}

A general, non-fundamental, dominant graph is obtained by taking one of the fundamental vacuum graph from Fig. \ref{vacuumLO} and performing repeated insertions on some propagator and in the prismatic bubble. The allowed propagator-insertions are listed on Fig. \ref{fig:covariance insertions} and the melonic insertion in the prismatic bubble is shown on Fig. \ref{fig:vertex insertion}. \\

\section{Conclusion}

We implemented the large-$N$ limit mechanism for the $O(N)^3$-invariant sextic tensor 
model, taking into consideration all $8$ invariant bubbles with optimal scalings. 

Using the intermediate field method and melonic moves, we reduced the 
initial set of sextic bubbles to a minimal set of building blocks, consisting 
only of the wheel and the tetrahedron bubbles:  
\begin{equation}
    \{I_b(T)\}_{b=1}^8 
    \xlongrightarrow{\text{intermediate field}} 
    \{I_2(T),~I_t(T,\chi),~I_p(T,\chi)\} 
    \xlongrightarrow{\text{Melonic moves}} 
    \{I_2(T), I_t(T,\chi)\}.
\end{equation}

This reduced model allowed us to identify dominant graphs. Planarity 
constraints on ribbon jackets ruled out fundamental dominant graphs mixing the wheel 
with other sextic interactions. Using the known results for the quartic case, we reconstructed the 
sextic representation, explicitly identifying the dominant graphs
and the degree conserving moves.

\begin{figure}[H]
    \centering
    \includegraphics[scale=0.5]{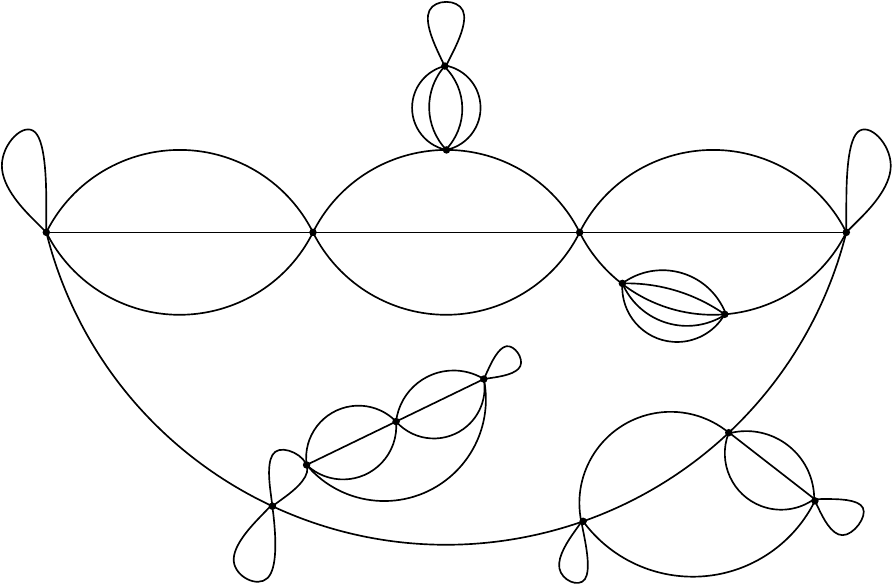}
    \caption{Example of a high-order dominant graph. Mixed structures together 
    with melonic insertions in the prismatic bubble generate intricate 
    diagrams.}
    \label{fig:uglyVac}
\end{figure}

\begin{figure}[H]
    \centering
    \includegraphics[scale=0.6]{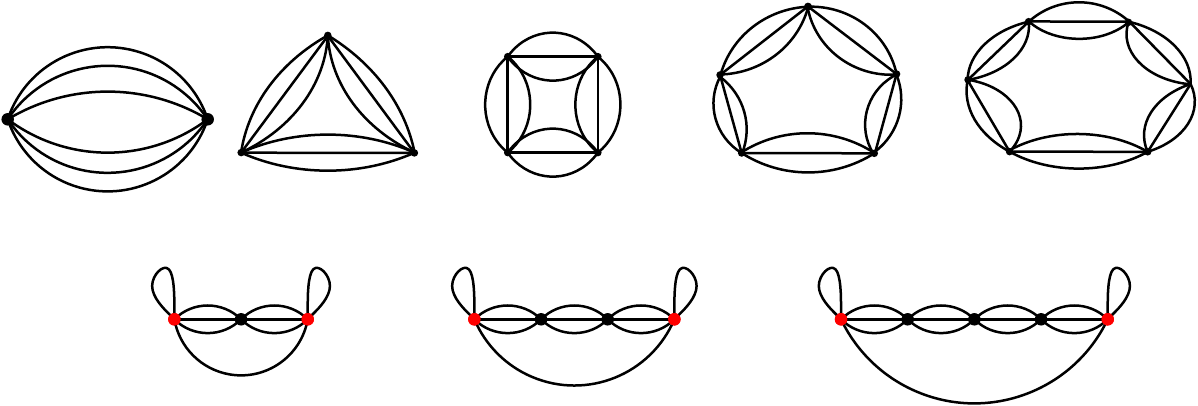}
    \caption{First graphs in the two infinite families generated by recursive 
    melonic insertions in the prismatic vertex. Edges denote the full $2-$point 
    function. Black dots are prismatic vertices; red dots are $g_5$ or $g_6$.}
    \label{fig:infiniteFamilies}
\end{figure}

The resulting large-$N$ structure is much richer than in previously 
studied tensor models. In particular, mixed-bubble structures and melonic 
insertions in the prismatic vertex generate complex dominant diagrams, as 
illustrated in Fig.~\ref{fig:uglyVac}, and even infinite families of dominant graphs, 
see Fig.~\ref{fig:infiniteFamilies}.  

Let us end this paper by listing 
several natural perspectives for future work:  
\begin{itemize}
    \item 
    Can the method developed here be 
    generalized to higher order interactions, to identify the minimal 
    set of bubbles from which all higher-order interactions can be reconstructed?  
    \item 
    The prismatic vertex insertion
    generates infinite sequences of dominant Feynman graphs. Is there a general criterion which
    determines which tensor interactions leads to such infinite families? Moreover, one can wonder whether or not higher-order bubbles can play a similar role?  
    \item 
    While melonic Feynman integrals are 
    well understood, the present model produces more intricate dominant graphs. Can 
    the corresponding Feynman integrals be classified and computed systematically, 
    potentially enabling progress on solving Schwinger--Dyson equations beyond the 
    melonic sector?  
    \item If one considers the model studied here as a a tensorial field theoretical model, non-trivial renormalization flow questions arise. This constitutes the topic of a companion paper \cite{future}.
\end{itemize}


%
%

\appendix



\section{Optimal scalings}
\label{optimal scalings}
\begin{figure}[H]
    \centering
    \includegraphics[scale=0.8]{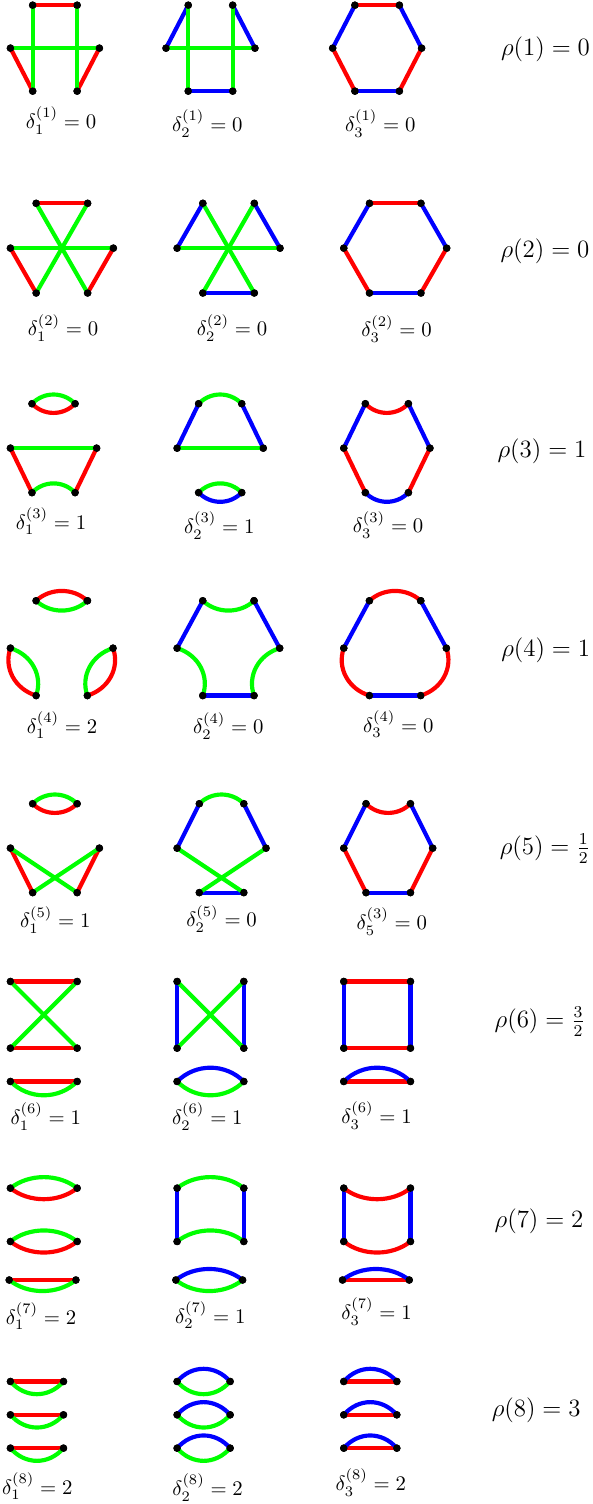}
    \caption{Optimal scalings computation}
    \label{fig:optimal scaling}
\end{figure}

The optimal scaling for interaction $b$ is $\rho(b)=\frac{1}{2}\sum_{l=1}^3\delta_l^{(b)}$, where $\delta_l^{(b)}=|J_l^{(b)}|-1$, with $|J_l^{(b)}|$ the number of connected components of the $l-$th jacket of the $b$-th bubble. \\
We show the three jackets for each bubble and scalings computation on Fig. \ref{fig:optimal scaling}.

\newpage
\bibliographystyle{ieeetr} 
\bibliography{main.bib}

@article{Giombi_2017,
   title={Bosonic tensor models at large {$N$} and small $\varepsilon$},
   volume={96},
   ISSN={2470-0029},
   url={http://dx.doi.org/10.1103/PhysRevD.96.106014},
   DOI={10.1103/physrevd.96.106014},
   number={10},
   journal={Physical Review D},
   publisher={American Physical Society (APS)},
   author={Giombi, Simone and Klebanov, Igor R. and Tarnopolsky, Grigory},
   year={2017},
   month=nov }

@article{Giombi_2018,
   title={Prismatic large {$N$} models for bosonic tensors},
   volume={98},
   ISSN={2470-0029},
   url={http://dx.doi.org/10.1103/PhysRevD.98.105005},
   DOI={10.1103/physrevd.98.105005},
   number={10},
   journal={Physical Review D},
   publisher={American Physical Society (APS)},
   author={Giombi, Simone and Klebanov, Igor R. and Popov, Fedor and Prakash, Shiroman and Tarnopolsky, Grigory},
   year={2018},
   month=nov }

@article{Benedetti_2020,
   title={Sextic tensor field theories in rank 3 and 5},
   volume={2020},
   ISSN={1029-8479},
   url={http://dx.doi.org/10.1007/JHEP06(2020)065},
   DOI={10.1007/jhep06(2020)065},
   number={6},
   journal={Journal of High Energy Physics},
   publisher={Springer Science and Business Media LLC},
   author={Benedetti, Dario and Delporte, Nicolas and Harribey, Sabine and Sinha, Ritam},
   year={2020},
   month=jun }

@article{Benedetti_2019,
   title={Line of fixed points in a bosonic tensor model},
   volume={2019},
   ISSN={1029-8479},
   url={http://dx.doi.org/10.1007/JHEP06(2019)053},
   DOI={10.1007/jhep06(2019)053},
   number={6},
   journal={Journal of High Energy Physics},
   publisher={Springer Science and Business Media LLC},
   author={Benedetti, Dario and Gurau, Razvan and Harribey, Sabine},
   year={2019},
   month=jun }

@article{jepsen2023rg,
    author = "Jepsen, Christian and Oz, Yaron",
    title = "{{RG} flows and fixed points of~{$O(N)^r$} models}",
    eprint = "2311.09039",
    archivePrefix = "arXiv",
    primaryClass = "hep-th",
    doi = "10.1007/JHEP02(2024)035",
    journal = "JHEP",
    volume = "02",
    pages = "035",
    year = "2024"
}

@article{Fraser-Taliente:2024rql,
    author = "Fraser-Taliente, Ludo and Wheater, John",
    title = "{Melonic limits of the quartic Yukawa model and general features of melonic CFTs}",
    eprint = "2410.09152",
    archivePrefix = "arXiv",
    primaryClass = "hep-th",
    doi = "10.1007/JHEP01(2025)187",
    journal = "JHEP",
    volume = "01",
    pages = "187",
    year = "2025"
}

@article{Krajewski_2023,
   title={Double scaling limit of the prismatic tensor model},
   volume={56},
   ISSN={1751-8121},
   url={http://dx.doi.org/10.1088/1751-8121/accf4e},
   DOI={10.1088/1751-8121/accf4e},
   number={23},
   journal={Journal of Physics A: Mathematical and Theoretical},
   publisher={IOP Publishing},
   author={Krajewski, T and Muller, T and Tanasa, A},
   year={2023},
   month=may, pages={235401} }

@article{future,
   title={Renormalization group flow of the {$O(N)^3-$}invariant general sextic tensor model},
     journal={to be submitted},
      author={Bardy, G and Krajewski, T and Muller, T and Tanasa, A},
   year={2025},
    }

@phdthesis{TheseTM,
  title={Study of combinatorial objects in higher dimensions},
  author={Muller, Thomas},
  year={2025},
  school={Universit{\'e} de Bordeaux}
}

@article{Bonzom_2022,
   title={Double scaling limit for the {$O(N)^3$}-invariant tensor model},
   volume={55},
   ISSN={1751-8121},
   url={http://dx.doi.org/10.1088/1751-8121/ac4898},
   DOI={10.1088/1751-8121/ac4898},
   number={13},
   journal={Journal of Physics A: Mathematical and Theoretical},
   publisher={IOP Publishing},
   author={Bonzom, V and Nador, V and Tanasa, A},
   year={2022},
   month=mar, pages={135201} }

@article{Carrozza_2016,
  title={{$O(N)$} Random Tensor Models},
  volume={106},
  ISSN={1573-0530},
  url={http://dx.doi.org/10.1007/s11005-016-0879-x},
  DOI={10.1007/s11005-016-0879-x},
  number={11},
  journal={Letters in Mathematical Physics},
  publisher={Springer Science and Business Media LLC},
  author={Carrozza, Sylvain and Tanasa, Adrian},
  year={2016},
  month=aug, pages={1531–1559} }

@article{Gurau_2011,
   title={The {1/$N$} Expansion of Colored Tensor Models},
   volume={12},
   ISSN={1424-0661},
   url={http://dx.doi.org/10.1007/s00023-011-0101-8},
   DOI={10.1007/s00023-011-0101-8},
   number={5},
   journal={Annales Henri Poincaré},
   publisher={Springer Science and Business Media LLC},
   author={Gurau, Razvan},
   year={2011},
   month=mar, pages={829–847} }

@article{Carrozza:2018psc,
    author = "Carrozza, Sylvain and Pozsgay, Victor",
    title = "{SYK-like tensor quantum mechanics with $\mathrm{Sp}(N)$ symmetry}",
    eprint = "1809.07753",
    archivePrefix = "arXiv",
    primaryClass = "hep-th",
    doi = "10.1016/j.nuclphysb.2019.02.012",
    journal = "Nucl. Phys. B",
    volume = "941",
    pages = "28--52",
    year = "2019"
}

@article{bonzom2015colored,
  title={Colored triangulations of arbitrary dimensions are stuffed Walsh maps},
  author={Bonzom, Valentin and Lionni, Luca and Rivasseau, Vincent},
  journal={The Electronic Journal of Combinatorics},
  year={2017},
  Volume = {24},
  Issue = {1},
}

@article{LIONNI2019600,
title = {Multi-critical behaviour of 4-dimensional tensor models up to order 6},
journal = {Nuclear Physics B},
volume = {941},
pages = {600-635},
year = {2019},
issn = {0550-3213},
doi = {https://doi.org/10.1016/j.nuclphysb.2019.02.026},
url = {https://www.sciencedirect.com/science/article/pii/S0550321319300604},
author = {Luca Lionni and Johannes Thürigen},
}

@book{Lionni_2018,
   title={Colored Discrete Spaces: Higher Dimensional Combinatorial Maps and Quantum Gravity},
   ISBN={9783319960234},
   ISSN={2190-5061},
   url={http://dx.doi.org/10.1007/978-3-319-96023-4},
   DOI={10.1007/978-3-319-96023-4},
   journal={Springer Theses},
   publisher={Springer International Publishing},
   author={Lionni, Luca},
   year={2018} }

@article{gurau_complete_2012,
	title = {The complete 1/{$N$} expansion of colored tensor models in arbitrary dimension},
	volume = {13},
	issn = {1424-0637, 1424-0661},
	url = {http://arxiv.org/abs/1102.5759},
	doi = {10.1007/s00023-011-0118-z},
	language = {en},
	number = {3},
	urldate = {2025-09-04},
	journal = {Annales Henri Poincaré},
	author = {Gurau, Razvan},
	month = apr,
	year = {2012},
	note = {arXiv:1102.5759 [gr-qc]},
	pages = {399--423},
}

@article{ambjorn_three-dimensional_1991,
        title = {Three-{Dimensional} {Simplicial} {Quantum} {Gravity} and {Generalized} {Matrix} {Models}},
        volume = {6},
        issn = {0217-7323},
        url = {https://ui.adsabs.harvard.edu/abs/1991MPLA....6.1133A},
        doi = {10.1142/S0217732391001184},
        urldate = {2025-09-10},
        journal = {Modern Physics Letters A},
        author = {Ambjørn, Jan and Durhuus, Bergfinnur and Jónsson, Thórdur},
        month = jan,
        year = {1991},
        note = {ADS Bibcode: 1991MPLA....6.1133A},
        pages = {1133--1146},
}

@article{sasakura_tensor_1991,
        title = {Tensor {Model} for {Gravity} and {Orientability} of {Manifold}},
        volume = {6},
        doi = {10.1142/S0217732391003055},
        journal = {Modern Physics Letters A - MOD PHYS LETT A},
        author = {Sasakura, Naoki},
        month = sep,
        year = {1991},
        pages = {2613--2623},
}

@article{Prakash_2020,
   title={Melonic dominance in subchromatic sextic tensor models},
   volume={101},
   ISSN={2470-0029},
   url={http://dx.doi.org/10.1103/PhysRevD.101.126001},
   DOI={10.1103/physrevd.101.126001},
   number={12},
   journal={Physical Review D},
   publisher={American Physical Society (APS)},
   author={Prakash, Shiroman and Sinha, Ritam},
   year={2020},
   month=jun }

@article{Keppler_2023_2,
   title={Duality of {$O(N)$} and {$Sp(N)$} random tensor models: tensors with symmetries},
   volume={56},
   ISSN={1751-8121},
   url={http://dx.doi.org/10.1088/1751-8121/ad0af4},
   DOI={10.1088/1751-8121/ad0af4},
   number={49},
   journal={Journal of Physics A: Mathematical and Theoretical},
   publisher={IOP Publishing},
   author={Keppler, H and Krajewski, T and Muller, T and Tanasa, A},
   year={2023},
   month=nov, pages={495206} }

@article{gurau2022dualityorthogonalsymplecticrandom,
    author = "Keppler, Hannes and Muller, Thomas",
    title = "{Duality of orthogonal and symplectic random tensor models: general invariants}",
    eprint = "2304.03625",
    archivePrefix = "arXiv",
    primaryClass = "hep-th",
    doi = "10.1007/s11005-023-01706-7",
    journal = "Lett. Math. Phys.",
    volume = "113",
    number = "4",
    pages = "83",
    year = "2023"
}

@article{Keppler_2023_1,
   title={Duality of orthogonal and symplectic random tensor models: general invariants},
   volume={113},
   ISSN={1573-0530},
   url={http://dx.doi.org/10.1007/s11005-023-01706-7},
   DOI={10.1007/s11005-023-01706-7},
   number={4},
   journal={Letters in Mathematical Physics},
   publisher={Springer Science and Business Media LLC},
   author={Keppler, Hannes and Muller, Thomas},
   year={2023},
}

@article{harribey_sextic_2022,
	title = {Sextic tensor model in rank 3 at next-to-leading order},
	volume = {2022},
	issn = {1029-8479},
	url = {https://doi.org/10.1007/JHEP10(2022)037},
	doi = {10.1007/JHEP10(2022)037},
	language = {en},
	number = {10},
	urldate = {2025-07-07},
	journal = {Journal of High Energy Physics},
	author = {Harribey, Sabine},
	month = oct,
	year = {2022},
	pages = {37},
}

@article{Rivasseau_2013,
   title={How to Resum Feynman Graphs},
   volume={15},
   ISSN={1424-0661},
   url={http://dx.doi.org/10.1007/s00023-013-0299-8},
   DOI={10.1007/s00023-013-0299-8},
   number={11},
   journal={Annales Henri Poincaré},
   publisher={Springer Science and Business Media LLC},
   author={Rivasseau, Vincent and Wang, Zhituo},
   year={2013},
   month=dec, pages={2069–2083} }

@article{lionni2016intermediate,
    author = "Lionni, Luca and Rivasseau, Vincent",
    title = "{Intermediate Field Representation for Positive Matrix and Tensor Interactions}",
    eprint = "1609.05018",
    archivePrefix = "arXiv",
    primaryClass = "math-ph",
    doi = "10.1007/s00023-019-00833-z",
    journal = "Annales Henri Poincare",
    volume = "20",
    number = "10",
    pages = "3265--3311",
    year = "2019"
}

@ARTICLE{1957SPhD....2..416S,
       author = {{Stratonovich}, R.~L.},
        title = "{On a Method of Calculating Quantum Distribution Functions}",
      journal = {Soviet Physics Doklady},
         year = 1957,
        month = jul,
       volume = {2},
        pages = {416},
       adsurl = {https://ui.adsabs.harvard.edu/abs/1957SPhD....2..416S}
}

@article{PhysRevLett.3.77,
  title = {Calculation of Partition Functions},
  author = {Hubbard, J.},
  journal = {Phys. Rev. Lett.},
  volume = {3},
  issue = {2},
  pages = {77--78},
  numpages = {0},
  year = {1959},
  month = {Jul},
  publisher = {American Physical Society},
  doi = {10.1103/PhysRevLett.3.77},
  url = {https://link.aps.org/doi/10.1103/PhysRevLett.3.77}
}

@article{HOOFT1974461,
title = {A planar diagram theory for strong interactions},
journal = {Nuclear Physics B},
volume = {72},
number = {3},
pages = {461-473},
year = {1974},
issn = {0550-3213},
doi = {https://doi.org/10.1016/0550-3213(74)90154-0},
url = {https://www.sciencedirect.com/science/article/pii/0550321374901540},
author = {G.'t Hooft}
}

@misc{marino2005leshoucheslecturesmatrix,
  author       = {Marino, Marcos},
  title        = {Les {H}ouches Lectures on Matrix Models and Topological Strings},
  year         = {2004},
  eprint       = {hep-th/0410165},
  archivePrefix = {arXiv},
  primaryClass = {hep-th},
  note         = {arXiv:hep-th/0410165}
}

@misc{eynard2018randommatrices,
  author        = {Eynard, Bertrand and Kimura, Taro and Ribault, Sylvain},
  title         = {Random Matrices},
  year          = {2015},
  eprint        = {1510.04430},
  archivePrefix = {arXiv},
  primaryClass  = {math-ph},
  note          = {arXiv:1510.04430 [math-ph]}
}

@inproceedings{ginsparg1993lectures2dgravity2d,
    author = {Ginsparg, Paul H. and Moore, Gregory W.},
    title = "{Lectures on {$2D$} gravity and {$2D$} string theory}",
    booktitle = "{Theoretical Advanced Study Institute (TASI 92): From Black Holes and Strings to Particles}",
    eprint = "hep-th/9304011",
    archivePrefix = "arXiv",
    reportNumber = "YCTP-P23-92, LA-UR-92-3479",
    pages = "277--469",
    month = "10",
    year = "1993"
}

@ARTICLE{difrancesco20042dquantumgravitymatrix,
       author = {{Di Francesco}, P.},
        title = "{2D Quantum Gravity, Matrix Models and Graph Combinatorics}",
      journal = {arXiv e-prints},
         year = 2004,
        month = jun,
          eid = {math-ph/0406013},
        pages = {math-ph/0406013},
          doi = {10.48550/arXiv.math-ph/0406013},
archivePrefix = {arXiv},
       eprint = {math-ph/0406013},
 primaryClass = {math-ph},
       adsurl = {https://ui.adsabs.harvard.edu/abs/2004math.ph...6013D}
}

@misc{ginsparg1991matrix,
       author = {{Ginsparg}, P.},
        title = "{Matrix models of 2d gravity}",
 howpublished = {Presented at the Trieste Summer School, Trieste (Italy), 22-25 Jul. 1991},
         year = 1991,
        month = jan,
        pages = {22-25},
          doi = {10.48550/arXiv.hep-th/9112013},
archivePrefix = {arXiv},
       eprint = {hep-th/9112013},
 primaryClass = {hep-th},
       adsurl = {https://ui.adsabs.harvard.edu/abs/1991tss..rept...22G}
}

@article{Brezin,
    author = "Brezin, E. and Itzykson, C. and Parisi, G. and Zuber, J. B.",
    title = "{Planar Diagrams}",
    reportNumber = "SACLAY-DPH-T-77-126",
    doi = "10.1007/BF01614153",
    journal = "Commun. Math. Phys.",
    volume = "59",
    pages = "35",
    year = "1978"
}

@article{DAVID198545,
title = {Planar diagrams, two-dimensional lattice gravity and surface models},
journal = {Nuclear Physics B},
volume = {257},
pages = {45-58},
year = {1985},
issn = {0550-3213},
doi = {https://doi.org/10.1016/0550-3213(85)90335-9},
url = {https://www.sciencedirect.com/science/article/pii/0550321385903359},
author = {F. David}
}

@article{KAZAKOV1985295,
title = {Critical properties of randomly triangulated planar random surfaces},
journal = {Physics Letters B},
volume = {157},
number = {4},
pages = {295-300},
year = {1985},
issn = {0370-2693},
doi = {https://doi.org/10.1016/0370-2693(85)90669-0},
url = {https://www.sciencedirect.com/science/article/pii/0370269385906690},
author = {V.A. Kazakov and I.K. Kostov and A.A. Migdal}
}

@article{Sachdev_1993,
   title={Gapless spin-fluid ground state in a random quantum {H}eisenberg magnet},
   volume={70},
   ISSN={0031-9007},
   url={http://dx.doi.org/10.1103/PhysRevLett.70.3339},
   DOI={10.1103/physrevlett.70.3339},
   number={21},
   journal={Physical Review Letters},
   publisher={American Physical Society (APS)},
   author={Sachdev, Subir and Ye, Jinwu},
   year={1993},
   month=may, pages={3339–3342} }

@article{witten2016syklikemodeldisorder,
    author = {Witten, Edward},
    title = {An SYK-Like Model Without Disorder},
    eprint = {1610.09758},
    archivePrefix = "arXiv",
    primaryClass = "hep-th",
    doi = "10.1088/1751-8121/ab3752",
    journal = {J. Phys. A},
    volume = {52},
    number = {47},
    pages = {474002},
    year = {2019}
}

@misc{Kitaev_2015,
       author = {{Kitaev}, Alexei},
        title = "{A simple model of quantum holography (part 1)}",
 howpublished = {Kavli Institute for Theoretical Physics Program: Entanglement in Strongly-Correlated Quantum Matter},
         year = 2015,
        month = apr,
          eid = {2},
        pages = {2},
       adsurl = {https://ui.adsabs.harvard.edu/abs/2015escq.progE...2K}
}

@article{Krishnan_2017,
   title={Quantum chaos and holographic tensor models},
   volume={2017},
   ISSN={1029-8479},
   url={http://dx.doi.org/10.1007/JHEP03(2017)056},
   DOI={10.1007/jhep03(2017)056},
   number={3},
   journal={Journal of High Energy Physics},
   publisher={Springer Science and Business Media LLC},
   author={Krishnan, Chethan and Sanyal, Sambuddha and Subramanian, P. N. Bala},
   year={2017},
   month=mar }

@article{GURAU2017386,
title = {The complete 1/{$N$} expansion of a {SYK}–like tensor model},
journal = {Nuclear Physics B},
volume = {916},
pages = {386-401},
year = {2017},
issn = {0550-3213},
doi = {https://doi.org/10.1016/j.nuclphysb.2017.01.015},
url = {https://www.sciencedirect.com/science/article/pii/S0550321317300299},
author = {Razvan Gurau}
}

@book{10.1093/oso/9780192895493.001.0001,
    author = {Tanasa, Adrian},
    title = "{Combinatorial Physics: Combinatorics, Quantum Field Theory, and Quantum Gravity Models}",
    publisher = {Oxford University Press},
    year = {2021},
    month = {04},
    isbn = {9780192895493},
    doi = {10.1093/oso/9780192895493.001.0001},
    url = {https://doi.org/10.1093/oso/9780192895493.001.0001},
}

@book{10.1093/acprof:oso/9780198787938.001.0001,
    author = {Gurău, Răzvan Gheorghe},
    title = {Random Tensors},
    publisher = {Oxford University Press},
    year = {2016},
    month = {10},
    isbn = {9780198787938},
    doi = {10.1093/acprof:oso/9780198787938.001.0001},
    url = {https://doi.org/10.1093/acprof:oso/9780198787938.001.0001},
}

@article{Eyal_1996,
   title={The {$O(N)$} vector model in the large {$N$} limit revisited: multicritical points and double scaling limit},
   volume={470},
   ISSN={0550-3213},
   url={http://dx.doi.org/10.1016/0550-3213(96)00168-X},
   DOI={10.1016/0550-3213(96)00168-x},
   number={3},
   journal={Nuclear Physics B},
   publisher={Elsevier BV},
   author={Eyal, Galit and Moshe, Moshe and Nishigaki, Shinsuke and Zinn-Justin, Jean},
   year={1996},
   month=jul, pages={369–395} }

@article{Moshe_2003,
   title={Quantum field theory in the large {$N$} limit: a review},
   volume={385},
   ISSN={0370-1573},
   url={http://dx.doi.org/10.1016/S0370-1573(03)00263-1},
   DOI={10.1016/s0370-1573(03)00263-1},
   number={3–6},
   journal={Physics Reports},
   publisher={Elsevier BV},
   author={Moshe, Moshe and Zinn-Justin, Jean},
   year={2003},
   month=oct, pages={69–228} }

@article{mulase_duality_2003,
	title = {Duality of Orthogonal and Symplectic Matrix Integrals and Quaternionic Feynman Graphs},
	volume = {240},
	url = {http://arxiv.org/abs/math-ph/0206011},
	doi = {10.1007/s00220-003-0918-1},
	language = {en},
	number = {3},
	urldate = {2025-09-09},
	journal = {Communications in Mathematical Physics},
	author = {Mulase, Motohico and Waldron, Andrew},
	month = sep,
	year = {2003},
	note = {arXiv:math-ph/0206011},
	pages = {553--586}
	}

@article{mkrtchyan_equivalence_1981,
	title = {The equivalence of {$Sp(2N)$} and {$SO(-2N)$} gauge theories},
	volume = {105},
	issn = {0370-2693},
	url = {https://www.sciencedirect.com/science/article/pii/0370269381910157},
	doi = {https://doi.org/10.1016/0370-2693(81)91015-7},
	number = {2},
	journal = {Physics Letters B},
	author = {Mkrtchyan, R. L.},
	year = {1981},
	pages = {174--176}
}
\end{document}